\documentclass[aps,prl,reprint,groupedaddress]{revtex4-2}
\usepackage{dcolumn}
\usepackage{bm}
\usepackage[utf8]{inputenc}
\usepackage[T1]{fontenc}
\usepackage{booktabs, array, mathptmx, float, tabularx, booktabs, lipsum, amsmath,multirow,amssymb}
\usepackage{siunitx, xcolor}
\usepackage[version=4]{mhchem}
\usepackage{graphicx,amssymb}
\usepackage{subfigure}
\usepackage{indentfirst,latexsym,bm}
\usepackage{threeparttable}
\graphicspath{{figs/}{figsgaoerb/}}
\usepackage[colorlinks,linkcolor=blue,anchorcolor=blue,citecolor=blue]{hyperref}
\usepackage{hyperref}
\usepackage{amsfonts}
\usepackage{newtxtext,newtxmath}

\begin{document}
	
\title{Symmetry-Engineered  Multiple Bulk-Boundary Correspondences and Anomalous Modes in a Non-Hermitian  Creutz Ladder}
\author{Xin Li$^{1}$}
\email{To whom correspondence should be addressed. Email: lixin$_$physics@126.com}
\author{TongYi Li$^{1}$}
\author{JingYu Peng$^{1}$}
\author{Yu Wang$^{1}$}
\affiliation{$^{1}$Data science research center, Kunming University of Science and Technology, Kunming, 650093, China}
	
	\begin{abstract}
		The synergy between non-Hermiticity and topology makes the bulk-boundary correspondence (BBC) highly elusive. Here
		we study a non-Hermitian Creutz ladder  incorporating both gain-loss and nonreciprocity, 
		and construct multiple BBCs involving scale-free, normal, and anomalous skin modes, as well as topological zero-energy modes. We characterize the skin effect induced by the parity-time ($\mathcal{P}\mathcal{T}$) phase transition through an average winding number, thereby formulating a $\mathcal{P}\mathcal{T}$-related BBC. Furthermore, a hidden chiral symmetry ensures that topological phase transitions can be reliably detected via a $\mathbb{Z}_{2}$ invariant. Although the gain-loss breaks the standalone  $\mathcal{P}$ symmetry, it preserves the combined $\mathcal{P}\mathcal{T}$ symmetry, allowing a modified $\mathbb{Z}_{2}$ invariant to sustain the topological BBC.  Notably, sublattice symmetry facilitates the precise analytical determination of non-Bloch spectra. Leveraging this, we introduce a hybrid spectral winding that encodes the localization (or delocalization) characteristics of two counterintuitive bulk modes coexisting with normal skin modes, thus defining a non-Hermitian BBC. One of these modes exhibits exponential boundary accumulation in a direction opposite to the nonreciprocity, while the other manifests as an anomalous surge of Bloch-wave states within the nonreciprocal lattice. These results reveal a series of unexpected phenomena governed by underlying symmetry, significantly broadening  our fundamental understanding of the BBC mechanism in non-Hermitian topological systems.
	\end{abstract}

	\maketitle

	\section{INTRODUCTION}\label{sec.1}
	The bulk-boundary correspondence (BBC) stands as a cornerstone criterion in topological band theory, establishing a rigorous intrinsic link between the Bloch bulk topology defined under periodic boundary conditions (PBCs) and the robust boundary states protected under open boundary conditions (OBCs). However, the emergence of non-Hermiticity poses fundamental challenges to this long-standing paradigm: in a broad class of nonreciprocal systems, the spectra under PBCs and OBCs exhibit drastic deviations, accompanied by the macroscopic accumulation of bulk states at boundaries — a universal phenomenon termed the non-Hermitian skin effect (NHSE). Tremendous research efforts have been dedicated to developing the non-Bloch band theory, which is built upon a fundamental generalization of the conventional Brillouin zone (BZ), namely the generalized Brillouin zone (GBZ) \cite{PhysRevLett.121.026808,PhysRevB.100.035102,PhysRevLett.123.066404,PhysRevLett.125.226402}. Specifically, the BBC projects on non-Hermitian systems follow two distinct lines of exploration. One line focuses on developing novel bulk invariants to describe conventional topological boundary states (topological BBC) \cite{PhysRevLett.121.026808,PhysRevB.111.L041406,PhysRevLett.121.086803,PhysRevB.99.081103,song2019non}. The other line builds a direct mapping between the NHSE and nontrivial spectral topology  (non-Hermitian  BBC) \cite{PhysRevB.109.035131,PhysRevLett.124.086801,RevModPhys.93.015005,PhysRevX.8.031079,wang2024non}.

	More recently, a new type of NHSE, namely critical NHSE (CNHSE), has further complicated the current situation \cite{li2020critical}. CNHSE refers to the dramatic sensitivity of OBC spectra to system size, which was first uncovered in systems with competing nonreciprocal coupling channels. It typically hosts scale-free (SF) skin modes and induces a strong size-dependent behavior of topological zero modes \cite{rafi2022system}.  Subsequent studies have further revealed that such SF localization can also be triggered by non-Hermitian impurities or boundary defects \cite{li2023scale,guo2023accumulation,li2021impurity}.
	These exotic phenomena are highly feasible for experimental verification across a wide range of state-of-the-art platforms, including 
	RLC electrical circuits \cite{helbig2020generalized}, acoustic ring resonators  \cite{gu2022transient}, optical fiber loops \cite{weidemann2020topological}, and coupled
	resonant optical waveguides \cite{zhu2020photonic}.

At its core, the rich physics of non-Hermitian systems is deeply intertwined with internal and crystalline symmetries, which  enrich the K-theory classification \cite{PhysRevX.8.031079,PhysRevB.99.235112,liu2019topological}, clarify the intrinsic features of the NHSE \cite{zhang2022review,PhysRevLett.124.086801,yi2020non,PhysRevB.103.205205}, and establish new paradigms for topological indices   \cite{liang2013topological,yin2018geometrical,zhang2019partial,brzezicki2019hidden,PhysRevLett.121.086803,PhysRevLett.121.086803,song2019non}.
  Of special interest is the parity-time ($\mathcal{P}\mathcal{T}$) symmetry, because a real-to-complex energy conversion across exceptional points (EPs) has given rise to  extraordinary applications such as EP-enhanced sensors  \cite{chen2017exceptional, hokmabadi2019non}, single-mode lasers \cite{hodaei2014parity}, optical isolators \cite{nazari2014optical} and unidirectional acoustic cloaks \cite{zhu2014pt}. However, research on $\mathcal{P}\mathcal{T}$-symmetric nonreciprocal topological systems is still limited, and the constraints imposed by $\mathcal{P}\mathcal{T}$ symmetry on the (C)NHSE warrant further clarification. Revealing the intricate interplay among symmetry, topology and non-Hermiticity will also lay a critical foundation for engineering novel device functionalities. 
  
  While prior studies have extensively elucidated the distinct BBC mechanisms that belong to either the topological or the non-Hermitian category, the coexistence of multiple BBCs in non-Hermitian systems endowed with diverse symmetries remains largely unexplored. Furthermore, whether the (C)NHSE inevitably disrupts the topological BBC constitutes an open question. To address this, we propose a non-Hermitian Creutz ladder that integrates balanced gain-loss and nonreciprocal hopping, wherein the $\mathcal{P}\mathcal{T}$ (pseudo-inversion) symmetry inherently encompasses a hidden chiral (sublattice) symmetry. Guided by the exact consistency of $\mathcal{P}\mathcal{T}$ symmetry breaking signatures between PBC and OBC spectra, we define an average winding number (AWN) to identify the emergence of  $\mathcal{P}$-symmetric (asymmetric) SF (exponential) skin modes. Beyond that, the hidden chiral symmetry endows the system with a well-defined $\mathbb{Z}_{2}$ topological invariant, which pinpoints the emergence of robust topological edge states. Benefiting from the exact non-Bloch spectral solvability guaranteed by sublattice symmetry, we find that the intersection and overlap of OBC spectra across different bulk mode families necessitate the construction of a hybrid spectral winding number. This new quantity generalizes the previously reported non-Hermitian BBC frameworks — which describe scenarios of PBC bands with opposite winding orientations or self-intersecting bands \cite{zeng2022non,xiao2024coexistence,wang2024non} — to the unexplored regime of degenerate PBC spectral loops. As a result,
the concurrent BBCs in a single non-Hermitian lattice give rise to topologically distinct boundary responses, each governed by its own unique localization mechanism.

	\section{model  and main arguments}\label{sec.2}

	Let us consider a non-Hermitian variant of the Creutz (cross-stitch) ladder with $N$ unit cells as sketched in Fig.~\ref{fig1} (a), for which the real-space Hamiltonian is given by 
	\begin{equation}\label{1}
	\begin{aligned}
	&H=\Sigma_{j=1}^{N}(J^{+}_{a}a_{j}^{\dag}a_{j+1}+J^{-}_{a}a_{j+1}^{\dag}a_{j}+J^{+}_{b}b_{j}^{\dag}b_{j+1}+J^{-}_{b}b_{j+1}^{\dag}b_{j})+\\
	&\Sigma_{j=1}^{N}(t_{1}a_{j+1}^{\dag}b_{j}+t_{2}b_{j+1}^{\dag}a_{j}+h.c.)+i\gamma\Sigma_{j=1}^{N}(a_{j}^{\dag}a_{j}-b_{j}^{\dag}b_{j}),\\
	\end{aligned}
	\end{equation}
	where  $a(b)^{\dag}_{j}$ is the creation operator at site $j$ acting on the upper (lower) leg of the ladder. In the diagonal bonds, nonuniform intercell hoppings $t_{1}=t_{0}(1-\delta)$ and $t_{2}=t_{0}(1+\delta)$ 
	with an unbalanced degree of $\delta\in[-1,1]$ are reciprocal, while the two legs own different nonreciprocal hoppings $J^{\pm}_{a(b)}=J[1\pm\eta_{a(b)}]$ with $\eta_{a(b)}\in[-1,1]$ and opposite gain-loss potentials $\pm i\gamma$. By adopting a vanishingly small interchain coupling, the CNHSE was first verified in 
	Ref.\cite{li2020critical}. Here, we lift  the  constraint of $t_{0}\ll1$ and instead set  $t_{0}=1$ as the energy unit throughout this paper. The core results derived under distinct Hamiltonian symmetries are presented in what follows:
	
	(1) Sec.~\ref{sec.3} (parameter regime: $\eta_{a}=-\eta_{b}$ and $\gamma=0$): This regime hosts independent, uncombined 
$\mathcal{P}$ and $\mathcal{T}$  symmetries, whose full implications on the Hamiltonian are systematically analyzed. The AWN and phase vortices characterize the $\mathcal{P}\mathcal{T}$ phase transition, which directly triggers the partial CNHSE. By virtue of a $\mathbb{Z}_{2}$ invariant protected by a hidden chiral symmetry,  we identify a valid case of the topological BBC for the $AI(R^{+})$ class of non-Hermitian systems that lack chiral symmetry but possess independent $\mathcal{P}$ and $\mathcal{T}$ symmetries \cite{liu2019topological}.

(2) Sec.~\ref{sec.4} (parameter regime: $\eta_{a}=-\eta_{b}$ and $\gamma\neq0$): Here, the system is governed by the combined global $\mathcal{P}\mathcal{T}$ symmetry. Breaking the $\mathcal{P}\mathcal{T}$ symmetry induces a spatially symmetric, bipolar NHSE. Remarkably, a corrected Bloch $\mathbb{Z}_{2}$ invariant remains capable of accurately predicting topological edge states, challenging the conventional wisdom that the NHSE necessitates a non-Bloch topological BBC with the GBZ \cite{PhysRevLett.121.086803,song2019non}.

(3) Sec.~\ref{sec.5} (parameter regime: $\eta_{a}\neq-\eta_{b}$ and $\gamma=0$): This section  investigates the regime with preserved 
 $\mathcal{T}$ symmetry but no global $\mathcal{P}\mathcal{T}$ symmetry,
which further contains two special subcases: the system with pseudo-inversion symmetry at $\eta_{a}=\eta_{b}$ and  the sublattice-symmetric limit at $\eta_{a}=\eta_{b}$ and $\delta=0$. A hybrid spectral winding number under sublattice symmetry signals the direction of anomalous NHSE, which is opposite to the dominant hopping. The $\mathbb{Z}_{2}$ invariant ceases to be valid in this case, as the hidden chiral symmetry undergoes spontaneous breaking.

	\section{independent  $\mathcal{P}$ and $\mathcal{T}$ symmetries }\label{sec.3}

The Hamiltonian Eq.~(\ref{1}) with $\eta_{a}=-\eta_{b}=\eta$ and $\gamma=0$ hosts  $\mathcal{P}$ and  (comlex-conjugate-type \cite{kawabata2019symmetry}) $\mathcal{T}$ symmetry separately:  $\mathcal{P}H\mathcal{P}^{-1}=H$, $\mathcal{K}H\mathcal{K}^{-1}=H$, where the parity operator has $\mathcal{P}a(b)_{j}\mathcal{P}^{-1}=b(a)_{N+1-j}$  and $\mathcal{K}(\equiv\mathcal{T})$ stands for complex conjugation. We defer the discussion on 
$\mathcal{P}$ symmetry breaking to Secs.\ref{sec.4} and \ref{sec.5}. Within the current section, the first two subsections are dedicated to the detailed analysis of the Bloch Hamiltonian, while third and fourth subsections focus on the SF skin modes and zero-energy modes respectively.

	\subsection{Various PBC properties}
	
	Imposing periodic boundary condition and inserting the Fourier transformation, i.e.
	$a(b)_{j}=\frac{1}{\sqrt{N}}\sum_{k=1}^{N}a(b)_{k}e^{\frac{2i\pi jk}{N}}$, the Hamiltonian reduces to $H=\sum_{k}\Psi_{k}^{\dag}\mathcal{H}(k)\Psi_{k}$ with the spinor $\Psi_{k}^{\dag}=\begin{bmatrix}a_{k}^{\dag}& b_{k}^{\dag} \\ \end{bmatrix}$, and the Bloch
	Hamiltonian reads 
	\begin{equation}\label{2}
	\mathcal{H}(k)=h_{0}I+h_{x}\sigma_{x}+h_{y}\sigma_{y}+ih_{z}\sigma_{z},
	\end{equation}
	where $h_{0}(k)=2J\cos (k),\,\,h_{x}(k)=(t_{1}+t_{2})\cos (k)$,
	$h_{y}(k)=(t_{1}-t_{2})\sin (k)$, $h_{z}(k)=2J\eta\sin (k)$, and $\sigma_{i}$ ($I$) is the Pauli (identity) matrix. The energy dispersion
	is obtained as
	\begin{equation}\label{3}
	E^{p}_{\pm}(k)=h_{0}(k)\pm2\sqrt{(t_{0}^{2}-t_{0}^{2}\delta^{2}+J^{2}\eta^{2})\cos ^{2}(k)+(t_{0}^{2}\delta^{2}-J^{2}\eta^{2})}.
	\end{equation}

	\begin{figure}\label{fig1}
		\scalebox{0.25}[0.25]{\includegraphics{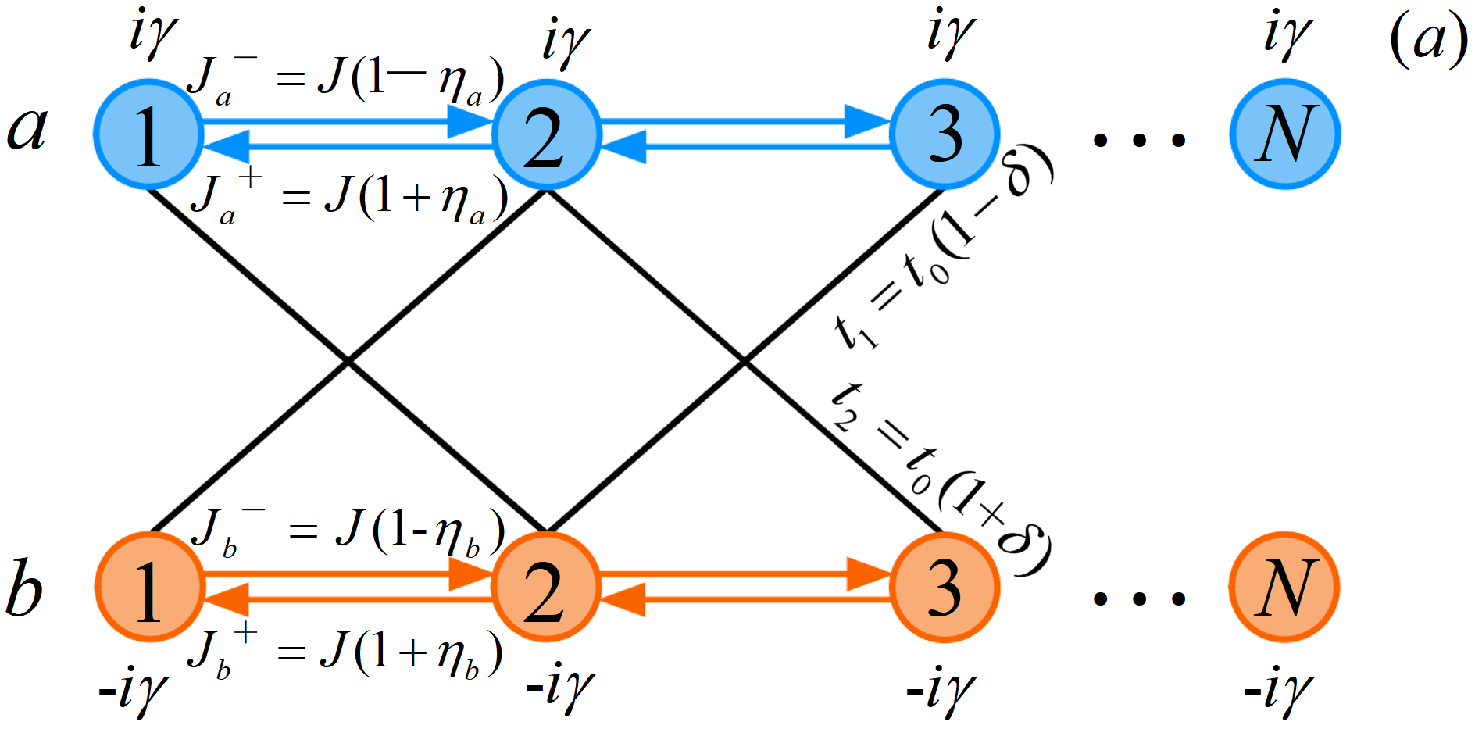}}	
		\scalebox{0.11}[0.11]{\includegraphics{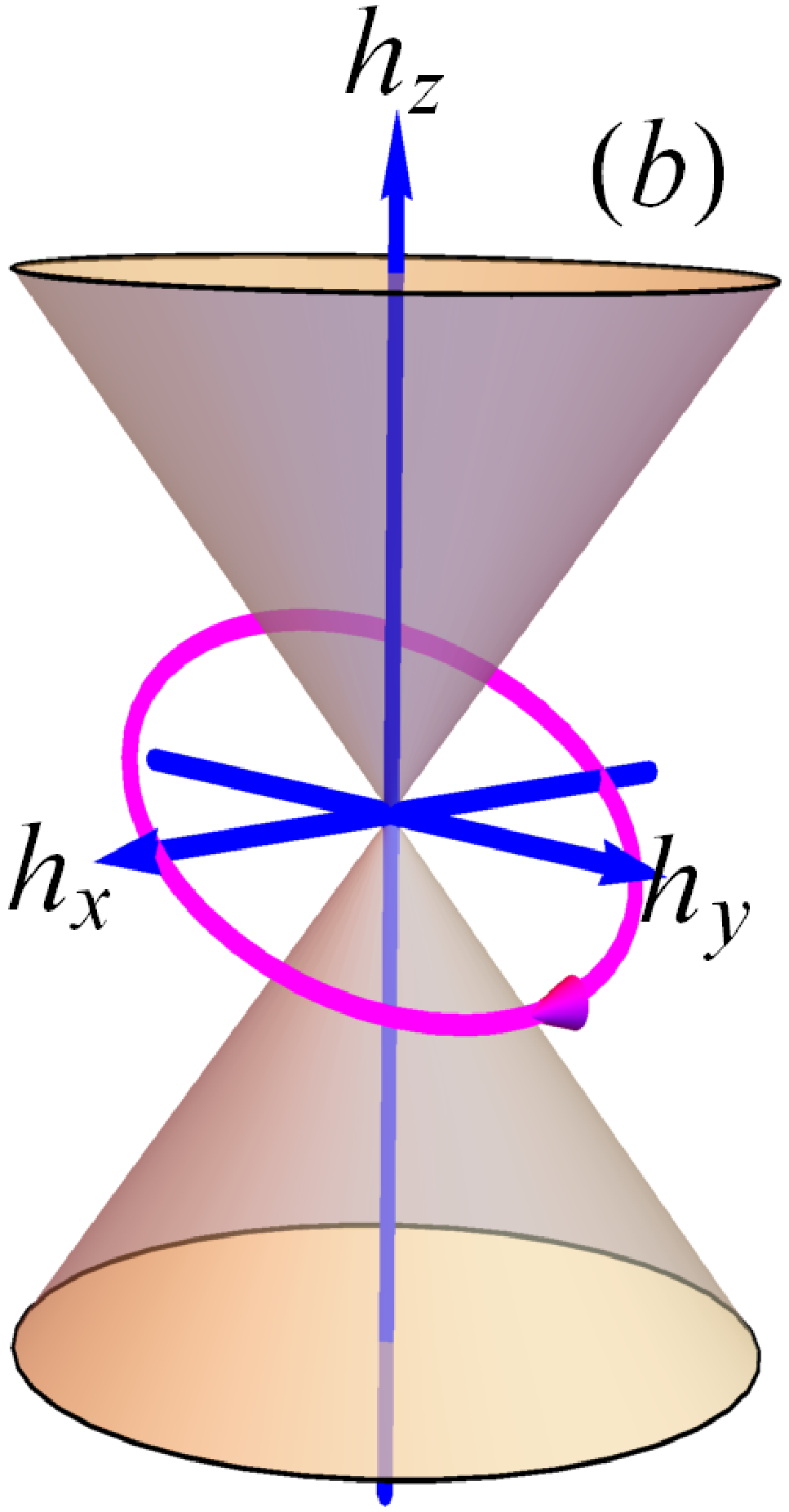}}
		\caption{\label{fig1}(color on line) (a) Schematic for the non-Hermitian ladder. (b) In the $\varGamma=-1$ zone of the phase diagram  Fig.~\ref{fig2} (a), the vector  $\bm{h}(k)=[h_{x}(k),h_{y}(k),h_{z}(k)]$ wraps clockwise around the conical surface $h^{2}_{x}+h^{2}_{y}=h^{2}_{z}$ once (in the top view).}
	\end{figure}

	We define left and right eigenstates of $\mathcal{H}(k)$ as:
	$\mathcal{H}(k)|\mathcal{R}_{\pm}(k)\rangle=E^{p}_{\pm}(k)|\mathcal{R}_{\pm}(k)\rangle$ and $\mathcal{H}^{\dag}(k)|\mathcal{L}_{\pm}(k)\rangle=E^{p\ast}_{\pm}(k)|\mathcal{L}_{\pm}(k)\rangle$ (``$\pm$'' indicate the band labels),  which become bi-orthogonal $\langle \mathcal{L}_{i}(k)|\mathcal{R}_{j}(k)\rangle=\delta_{ij}$. the $\mathcal{P}$ symmetry $\sigma_{x}\mathcal{H}(k)\sigma_{x}=\mathcal{H}(-k)$ ensures $\sigma_{x}|\mathcal{R}(\mathcal{L})_{\pm}(-k)\rangle\varpropto|\mathcal{R}(\mathcal{L})_{\pm}(k)\rangle$ with $E^{p}_{\pm}(k)=E^{p}_{\pm}(-k)$; meanwhile the $\mathcal{T}$ symmetry $\mathcal{T}\mathcal{H}(k)\mathcal{T}^{-1}=\mathcal{H}(-k)$  renders  $|\mathcal{R}^{\ast}(\mathcal{L}^{\ast})_{\pm}(k)\rangle\varpropto|\mathcal{R}(\mathcal{L})_{\pm}(-k)\rangle$ [$\varpropto|\mathcal{R}(\mathcal{L})_{\mp}(-k)\rangle$]  for real (complex) $E^{p}_{\pm}(k)$ because the complex bands are
	paired as $E^{p\ast}_{\pm}(k)=E^{p}_{\mp}(-k)$, but $|\mathcal{R}_{\pm}(k)\rangle\varpropto\sigma_{y}|\mathcal{L}^{\ast}_{\mp}(k)\rangle$ always holds. Further considering that $E^{p}_{+}(k)=-E^{p}_{-}(\pi-k)$, the spectrum 
	is symmetric about both the real and imaginary 
	axes. The $\mathcal{P}\mathcal{T}$ symmetry breaking is obviously aimed at $|\mathcal{R}(\mathcal{L})_{\pm}(k)\rangle$ rather than  $\mathcal{H}(k)$, the former ceases to be an eigenstate of $\mathcal{P}\mathcal{T}$ operator once $E^{p}_{\pm}(k)$ is complex.
	
\begin{figure}\label{fig2}
	\subfigure{
		\includegraphics[width=0.24\textwidth]{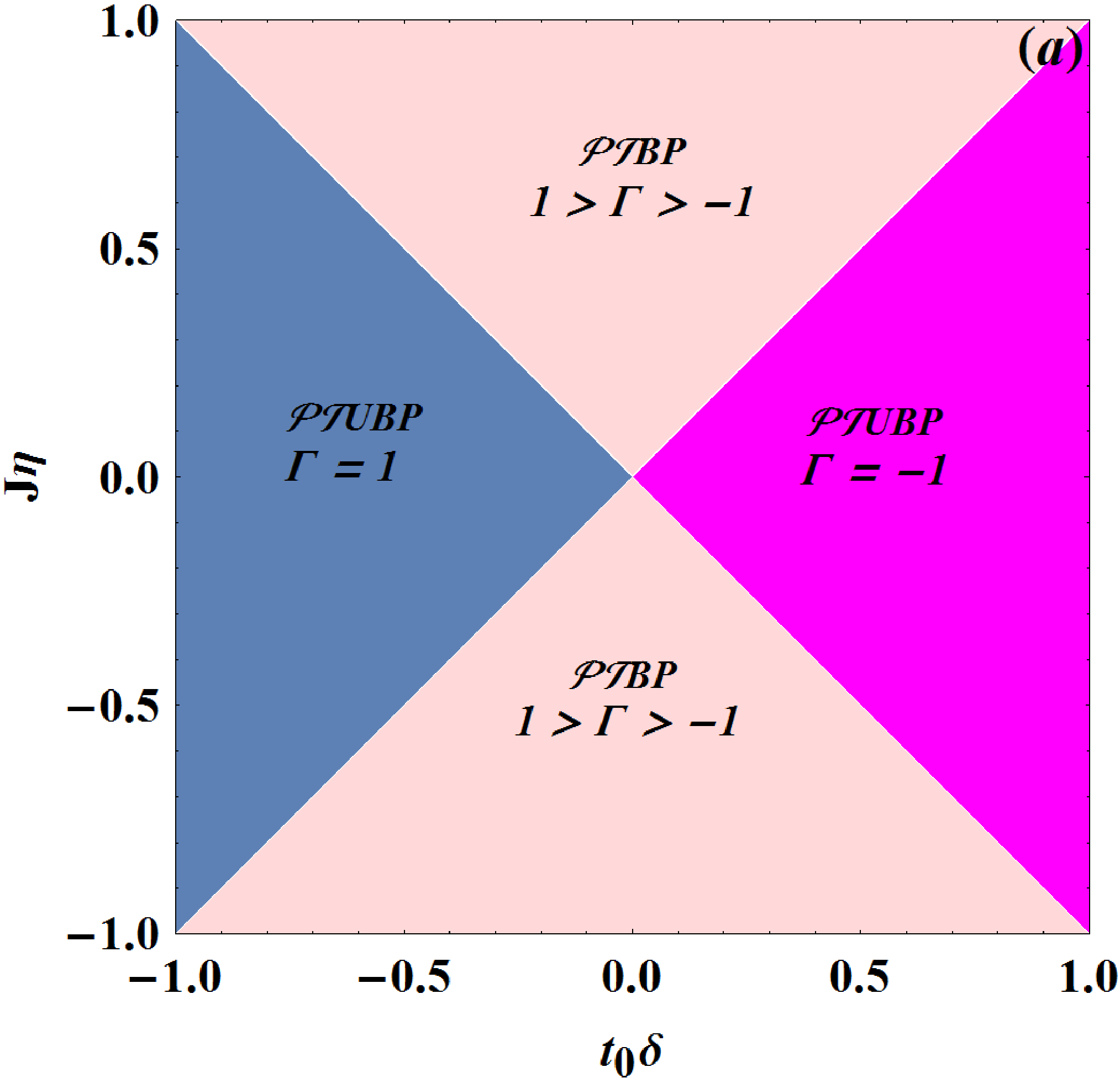}}
	\begin{minipage}[b]{0.215\textwidth} 
		
		\includegraphics[width=\textwidth]{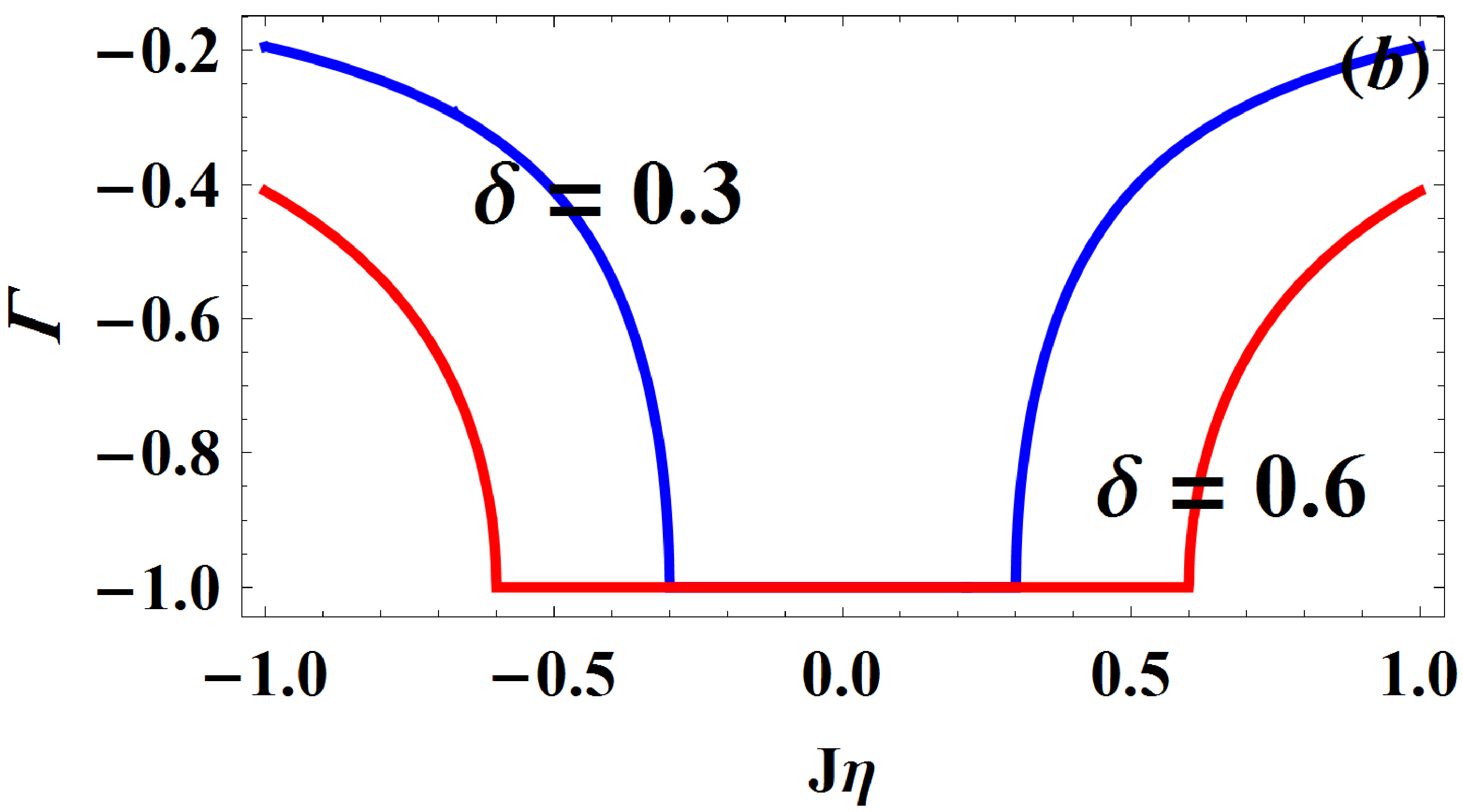}
		\includegraphics[width=\textwidth]{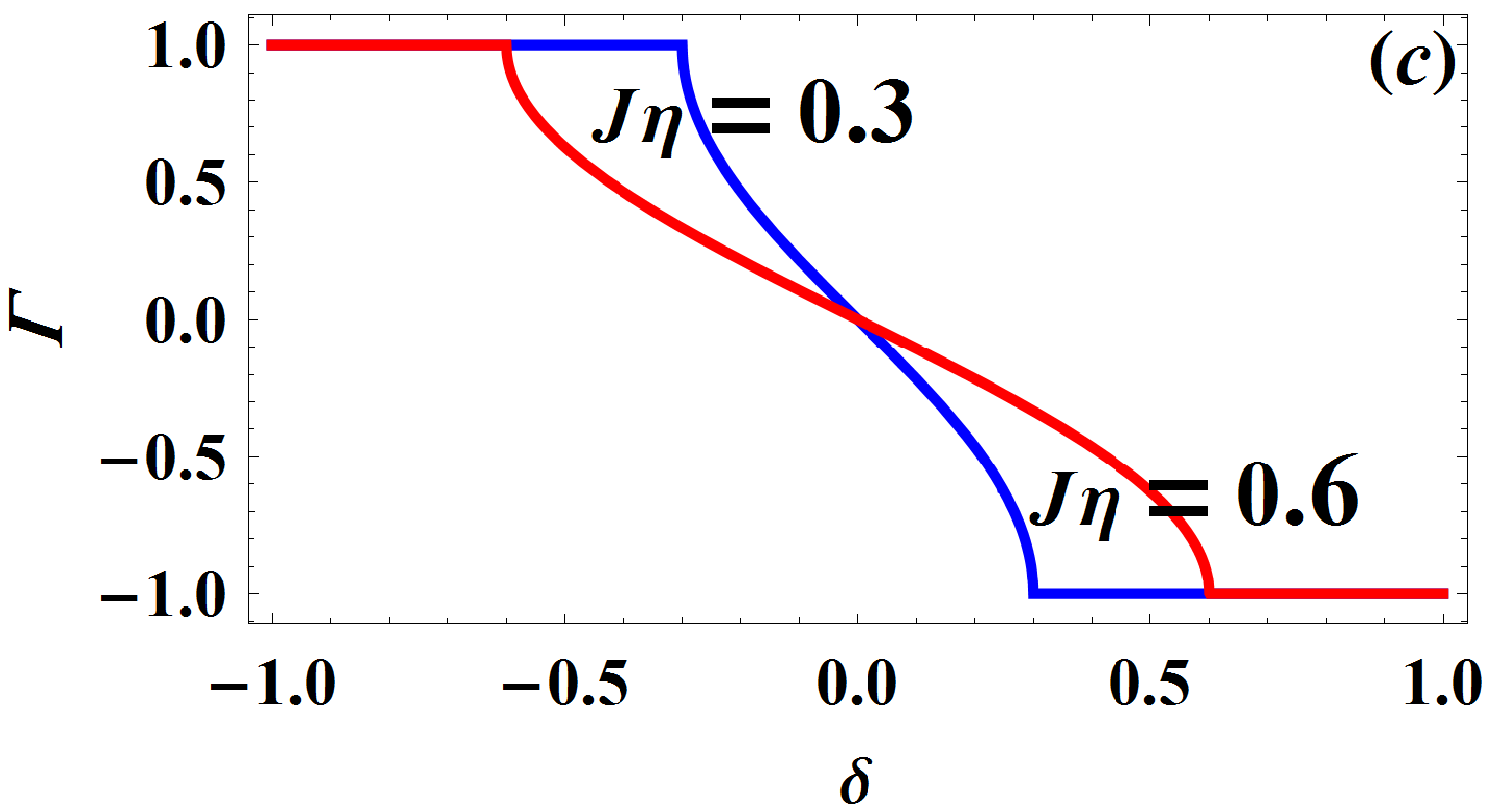}
	\end{minipage}
	\caption{\label{fig2}(color on line) (a) $\mathcal{P}\mathcal{T}$ phase diagram of the Hamiltonian Eq.~(\ref{2}) with the boundary  determined by $|J\eta|=|t_{0}\delta|$ in the $J\eta-t_{0}\delta$ plane: $1>\varGamma>-1$ (lightred), $\varGamma=1$(navyblue) and $\varGamma=-1$ (magenta). (b) $\varGamma$ versus $J\eta$ for $\delta=0.3$ (blue) and  $\delta=0.6$ (red). (c) $\varGamma$ versus $\delta$ for $J\eta=0.3$ (blue) and  $J\eta=0.6$ (red). Set $t_{0}=1$.}
\end{figure}	
	
	\subsection{AMN and vortices at exception points}

	The irremovable energy offset $h_{0}$, which  leaves the eigenstates unchanged, breaks chiral symmetry. This generally makes the conventional winding number $\mathcal{W}_{\pm}=\frac{1}{\pi}\int^{2\pi}_{0}$ $\langle \mathcal{L}_{\pm}|i\partial_{k}|\mathcal{R}_{\pm}\rangle dk$ non-integer, while their sum $\mathcal{W}=\mathcal{W}_{+}+\mathcal{W}_{-}$ stays integer-valued \cite{liang2013topological}. In sharp contrast, the periodic real-valued component $h_{z}(k)$ under our case of Eq.~(\ref{2})
	yields $\mathcal{W}_{+}=\mathcal{W}_{-}=\pm1$ for $\delta\lessgtr0$, indicating that
	the closed trajectory traced by the vector $\bm{h}(k)=[h_{x}(k),h_{y}(k),h_{z}(k)]$ winds once around the $h_{z}$-axis  in the $\bm{h}$-vector space. Nevertheless, this bulk topology does not induce significant changes in the eigenstate with OBC. As will be demonstrated later, all bulk eigenstates are extended in $\mathcal{P}\mathcal{T}$ UBP, whereas breaking the $\mathcal{P}\mathcal{T}$ symmetry triggers the emergence of a substantial number of SF skin modes. This phenomenon is intrinsically linked to the topology of the energy loop encircling the kissing cones $h_{x}^{2}+h_{y}^{2}=h_{z}^{2}$ under PBC. Since an azimuthal angle $\phi\in[-\pi,\pi)$ always exists such that $h_{x}-ih_{y}=h_{z}e^{-i\phi}$ at the EP, we invoke a Hermitian topological counterpart of $\mathcal{H}(k)$: 
	\begin{equation}\label{4}
	\begin{aligned}
	&\mathcal{H}^{\prime}(k,\phi)=h_{x}^{\prime}\sigma_{x}+h_{y}^{\prime}\sigma_{y},\\
	&h_{x}^{\prime}=h_{x}- h_{z}\cos(\phi),\,\,h_{y}^{\prime}=h_{y}- h_{z}\sin(\phi),\\
	\end{aligned}
	\end{equation}
	which can be realized by a long-range dimerized lattice incorporating hopping phase factors \cite{zheng2024phase,ahmadi2020topological}. 
	Given the restoration of chiral symmetry in $\mathcal{H}^{\prime}(k,\phi)$, and utilizing the eigenequation $\mathcal{H}^{\prime}(k,\phi)|\varphi_{\pm}\rangle=\pm\varepsilon|\varphi_{\pm}\rangle$ ($\langle\varphi_{\pm}|\varphi_{\pm}\rangle=1$), we can define an AWN:
	\begin{equation}\label{5}
	\begin{aligned}
	\varGamma=\int^{\pi}_{-\pi}\mathcal{W}_{\phi}\frac{d\phi}{2\pi}
	&\left\{
	\begin{aligned}
	=\pm \mathbb{Z}^{\ast}, \mathcal{P}\mathcal{T}UBP,\\
	\neq \pm \mathbb{Z}^{\ast},\,\, \,\,\,\,\,\,\,\mathcal{P}\mathcal{T}BP,
	\end{aligned}
	\right.\\
	\end{aligned}
	\end{equation}
	which elucidates the unique topological structure of the system. The non-zero integer
	$\mathbb{Z}^{\ast}$ quantifies the winding count around the conical surface of exceptional points (``$\pm$'' indicats direction) [Fig.~\ref{fig1} (b)]. Specifically, the winding numbers $\mathcal{W}_{\phi}=\frac{1}{\pi}\int^{2\pi}_{0}\langle \varphi_{\pm}|i\partial_{k}|\varphi_{\pm}\rangle dk$ stay identical for all $\phi$ provided that the fictitious Hamiltonian $\mathcal{H}^{\prime}(k,\phi)$ remains non-degenerate across the entire range of $\phi$. Otherwise, the loop traced by $\bm{h}(k)$ will pierce through the kissing cones.
	The phase diagram presented in Fig.~\ref{fig2} (a-c) demonstrate that the AWN serves as avalid topological invariant within $\mathcal{P}\mathcal{T}$-unbroken region $|J\eta|<|t_{0}\delta|$, whereas it becomes non-integer in the $\mathcal{P}\mathcal{T}$-broken region	$|J\eta|>|t_{0}\delta|$. Consequently, the $\mathcal{P}\mathcal{T}$ phase transition in the original non-Hermitian ladder is accurately predicted. This indicates that the bulk eigenstates in $\mathcal{P}\mathcal{T}$ UBP are protected by both $\mathcal{P}\mathcal{T}$ symmetry and topology. For a pair of Nelson-Hatano (NH) chains with only rung couplings,  the parameters in Eq.~(\ref{2}) reduce to $h_{x}=t_{0}$ and $h_{y}=0$ \cite{li2020critical,yokomizo2021scaling}.  In this configuration,  the inherent topology in $\mathcal{P}\mathcal{T}$ UBP is absent, but
	the AWN remains effective if the Berry connection $\langle \varphi_{\pm}|i\partial_{k}|\varphi_{\pm}\rangle$ is replaced by $\langle \varphi_{\pm}|i\partial_{\phi}|\varphi_{\pm}\rangle$. There we obtained $\varGamma=0(<0)$ in $\mathcal{P}\mathcal{T}$ UBP (BP).

	In the Hermitian limit ($\eta=0$), the band gap close and then reopen as $\delta\rightarrow-\delta$.  This phase transition admits a formal topological description in terms of the winding number $\mathcal{W}_{\phi}(\delta\lessgtr0,\eta=0)=\pm1$. Yet, it should be emphasized that, since the $h_{0}$ term breaks chiral symmetry, 
	the emergence of zero-energy edge states is entirely decoupled from $\mathcal{W}_{\phi}$, distinguishing this system from SSH models. Specifically, when  $t_{0}>J$, the edge states appear and the two bands become fully gapped (also referred to as `isolated' in Ref.\cite{shen2018topological}) i.e., $E^{p}_{+}(k_{+})\neq E^{p}_{-}( k_{-})$ for all $k_{\pm}$. Therefore, an alternative bulk invariant is required to recover the topological BBC; this will be addressed subsequently in the context of the non-
	Hermitian case.

	\begin{figure}\label{fig3}
		\scalebox{0.125}[0.125]{\includegraphics{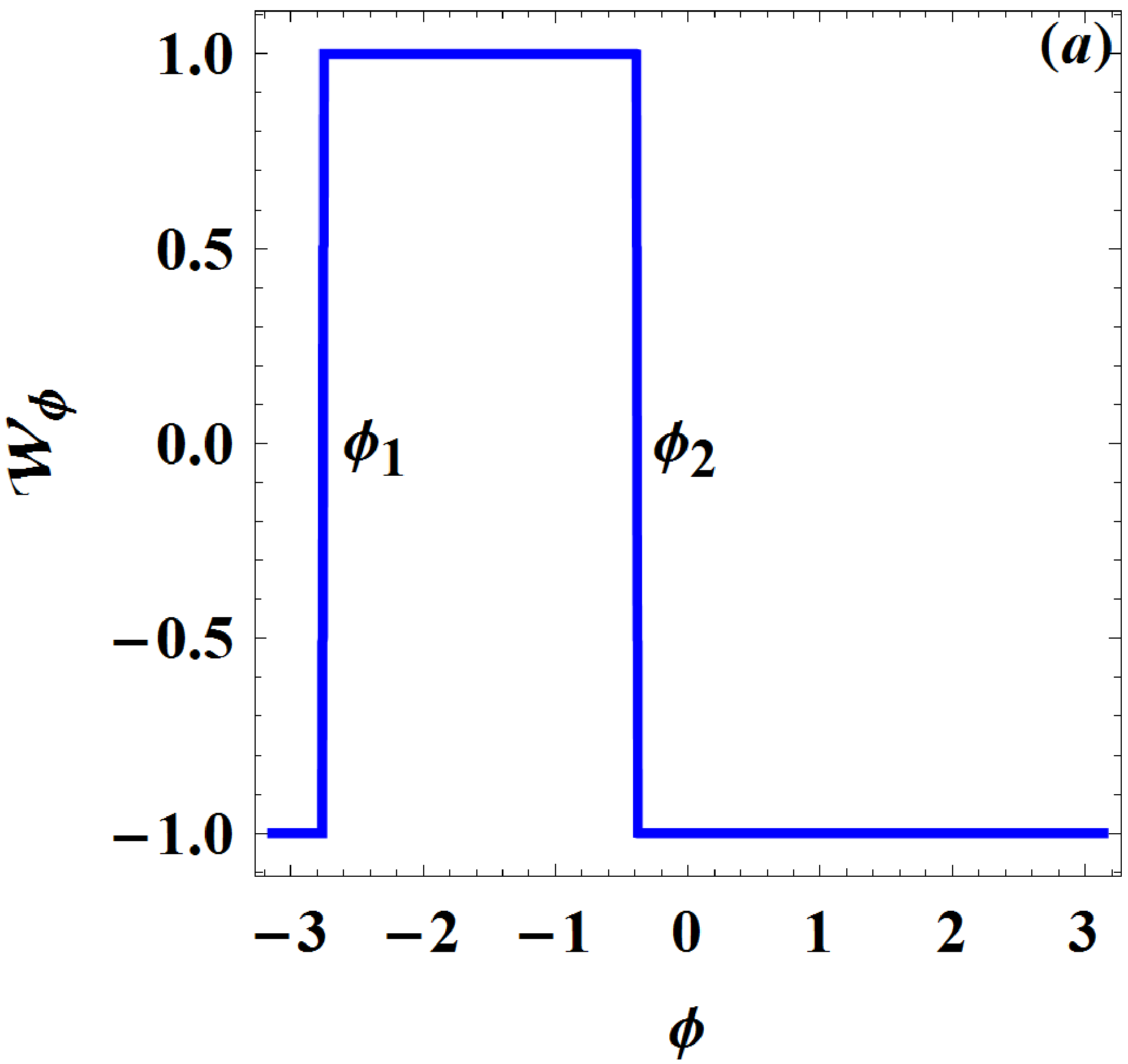}}	\scalebox{0.132}[0.132]{\includegraphics{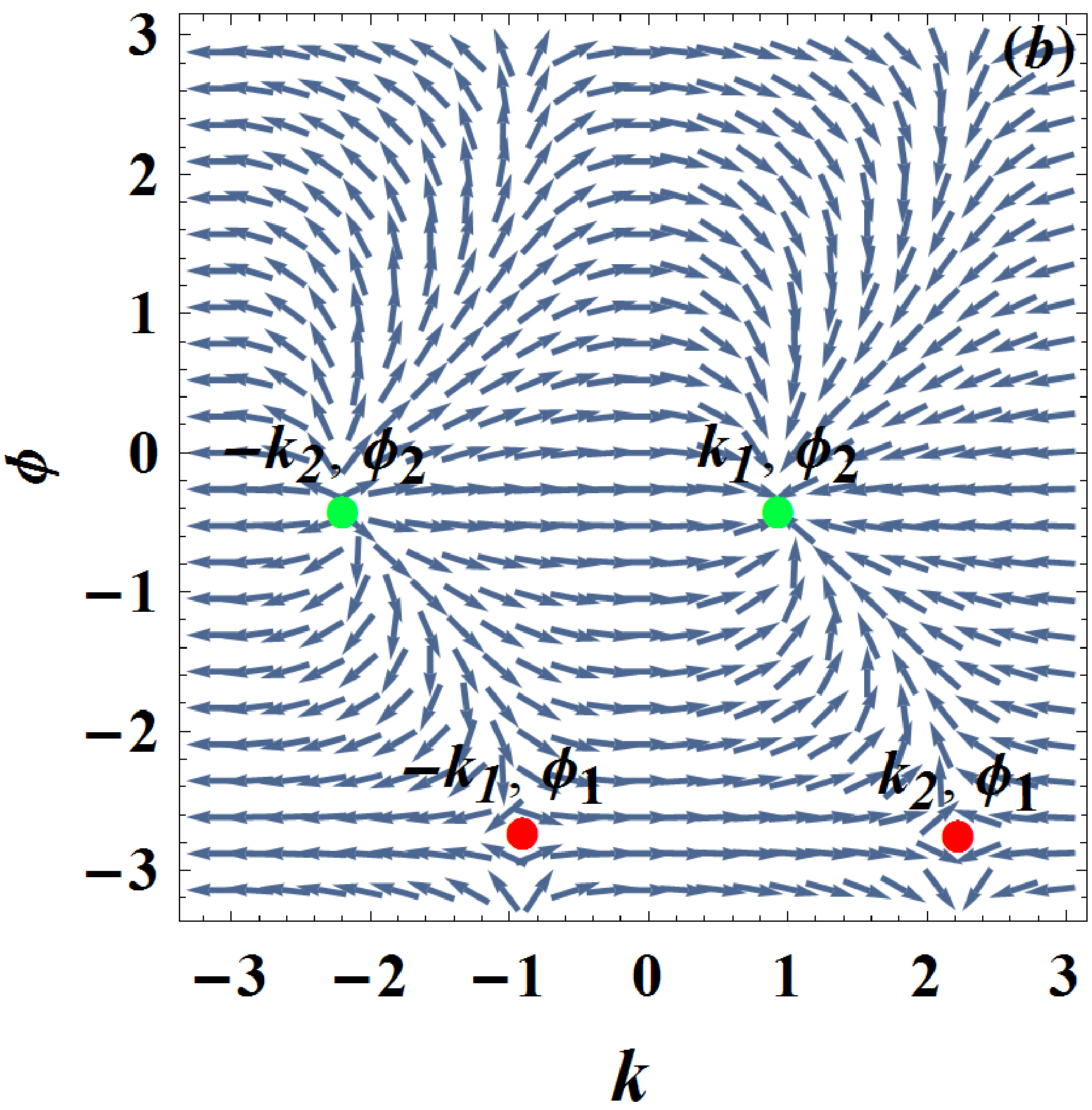}}
		\caption{\label{fig3}(color on line) (a) Flipping the winding number of $\mathcal{H}^{\prime}(k,\phi)$ in the interval $(\phi_{1},\phi_{2})$. (b)
			the red and green dots refer to vortices $(\mathcal{V}_{e}=1)$ and anti-vortices  $(\mathcal{V}_{e}=-1)$, respectively.
			$J\eta=0.8$, $\delta=0.3$, $k_{1}\approx 0.93$, $k_{2}\approx 2.21$, $\phi_{1}\approx-2.76$, and $\phi_{2}\approx-0.38$.}
	\end{figure}
	
	At the critical condition $|t_{0}\delta|=|J\eta|$, the two bands coalesce at two EPs $k=\pm\frac{\pi}{2}$. Concurrently, the winding number inherited from the Hermitian counterpart starts to
	reverse its sign at $\phi_{0}=-\frac{\pi}{2}$ ($\frac{\pi}{2}$) for $\eta>0$ ( $\eta<0$). When $|J\eta|>|t_{0}\delta|$, the two EPs bifurcate into four:  $\pm k_{1}=\pm\arccos\sqrt{\xi}$, and $\pm k_{2}=\pm(\pi-\arccos \sqrt{\xi})$, where  $\xi=(J^{2}\eta^{2}-t_{0}^{2}\delta^{2})/[t_{0}^{2}(1-\delta^{2})+J^{2}\eta^{2}]$.  As  $|J\eta|$ increases further, the complex energy regions
	$(k_{1},k_{2})$ and  $(-k_{2},-k_{1})$ expand, corresponding to a broader winding reversal zone $(\phi_{1},\phi_{2})$ for $\mathcal{H}^{\prime}(k,\phi)$ [see Fig.~\ref{fig3} (a)]. Because the four diabolic points \cite{arkhipov2023dynamically} $(k_{i},\phi_{j})$ of this Hermitian counterpart are precisely associated with EPs in the original non-Hermitian system,   
	the spin polarizations 
	$\pm\bm{F}(k,\phi)=\pm[F_{x}(k,\phi),F_{y}(k,\phi)]$($\equiv\langle\varphi_{\pm}|\sigma_{x}|\varphi_{\pm}\rangle,\langle\varphi_{\pm}|\sigma_{y}|\varphi_{\pm}\rangle$) can define a planar vector field
	in 2D pseudo-BZ $\bm{k}\equiv(k,\phi)$. Within this field, the emergence of vortices characterized by 
	\begin{equation}\label{6}
	\mathcal{V}_{e}=\oint_{C}\frac{d\bm{k}}{2\pi}(F_{x}\partial_{\bm{k}} F_{y}-F_{y}\partial_{\bm{k}} F_{x})
	\end{equation} 
	signifies the $\mathcal{P}\mathcal{T}$ phase transition.  The line integral in Eq.~(\ref{6}) is performed around diabolic points $(k_{i},\phi_{j})$, and the vortices are invariably generated and annihilated in pairs as illustrated in Fig.~\ref{fig3} (b). This synthetic topological defect realized via dimensional expansion
	plays analogous to that of Weyl nodes in a 2D fermionic square lattice \cite{hou2013hidden}.

	\subsection{$\mathcal{P}\mathcal{T}$-related BBC}

	\begin{figure}\label{fig4}
		\scalebox{0.157}[0.157]{\includegraphics{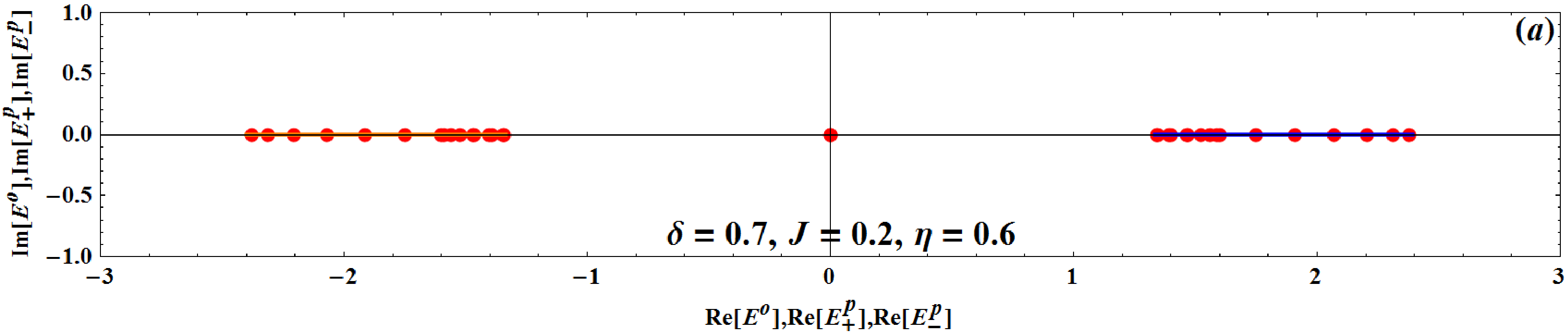}}
		\scalebox{0.149}[0.149]{\includegraphics{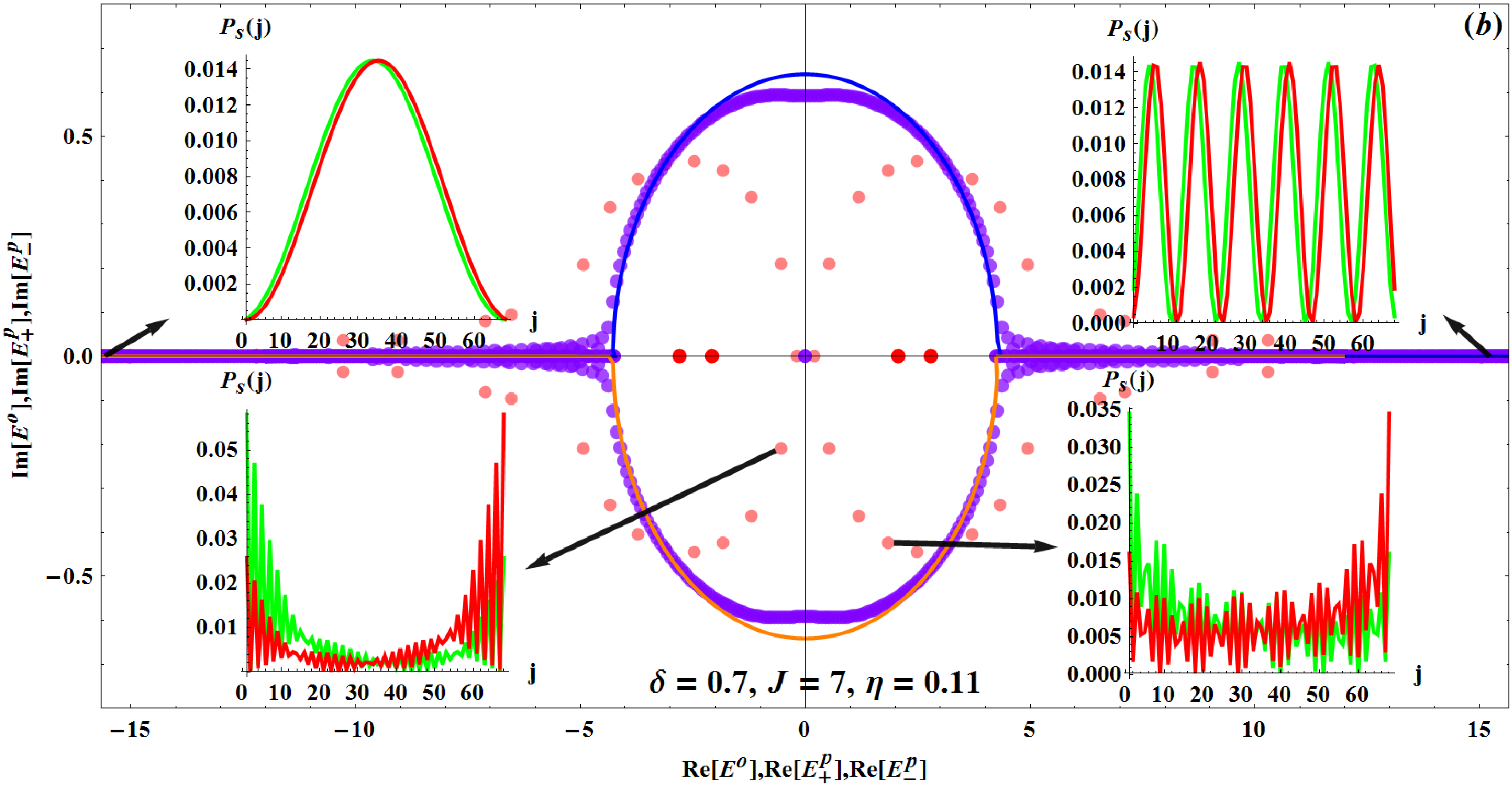}}	\scalebox{0.162}[0.162]{\includegraphics{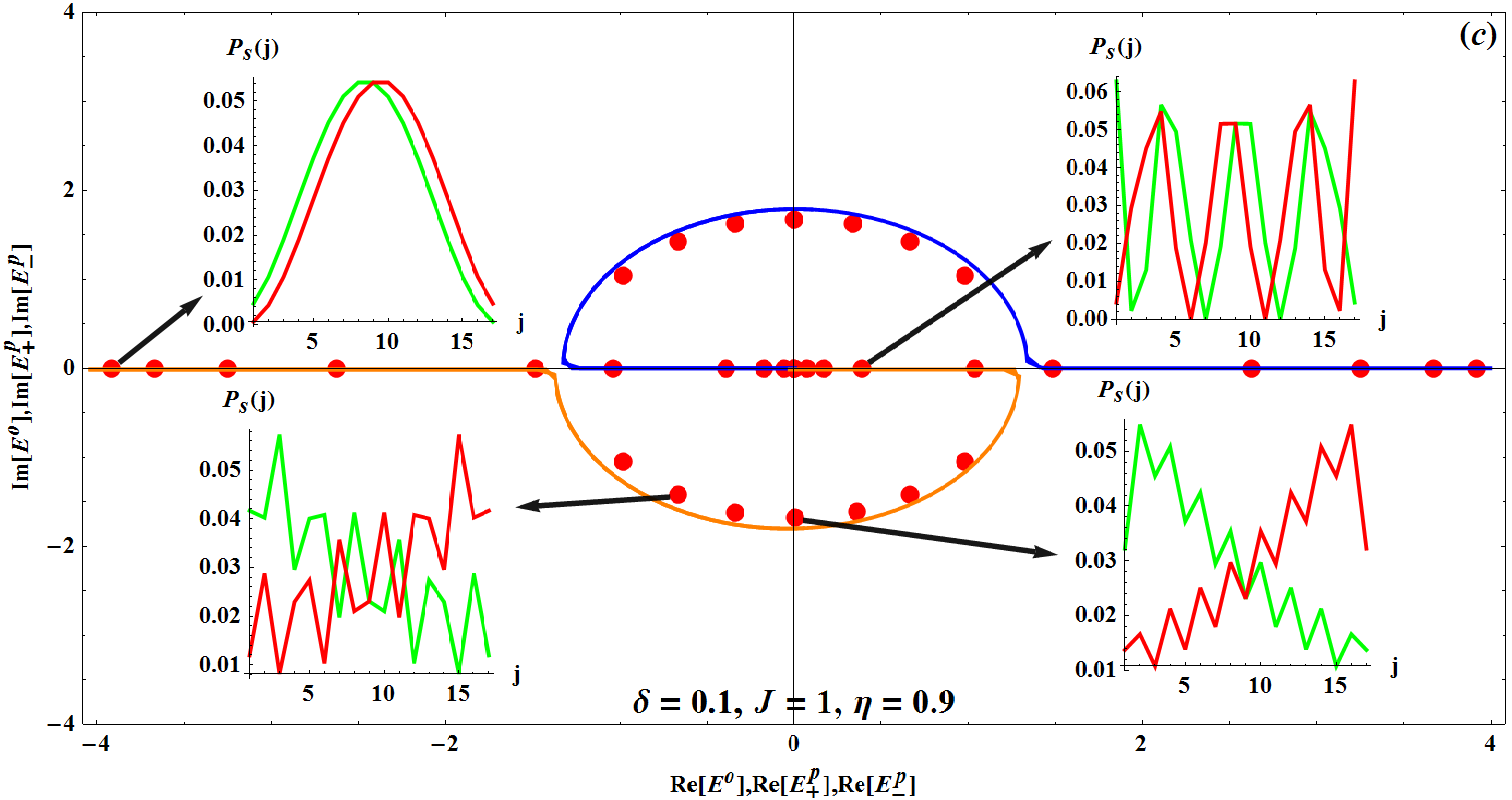}}
		
		\caption{\label{fig4}(color on line) PBC energy spectra	$E^{p}_{-}(k)$ (orange line), $E^{p}_{+}(k)$ (blue line) and OBC spectrum $E^{o}$ (dots) on the complex plane.  $E^{o}$ in (a,b) also includes the isolated zero-energy points corresponding to topological edge modes. Here, $N=17$ (red dots) in (a,c),  $N=8$ (red dots) in (b), $N=68$ (pink dots) in (b), $N=568$ (purple dots) in (b). The merging of energy bands just occurs in (c). Insets: the density profile $P_{s}(j)(s=a,b)$ as a function of site index $j$ for sub-sites $a$ (green) and  $b$ (red).}
	\end{figure}

	Fig.~\ref{fig4} (a) and (b, c) depict
	the energy spectra within the $\mathcal{P}\mathcal{T}$ UBP and BP, respectively. In the regime where $J\gg t_{0}$,  the OBC spectrum in $\mathcal{P}\mathcal{T}$ BP undergoes a gradual transition from the real axis to the PBC spectrum  as $N$ increases (CNHSE). Upon reaching a critical system size, two degenerate zero-energy modes emerge prominently within the point gap [isolated  purple dot in Fig.~\ref{fig4} (b)]. As discussed in Subsection~D, this size-dependent effect only emerges for even $N$. Conversely, smaller $J$ makes the OBC and PBC spectra match very well even in small-sized systems as demonstrated in Fig.~\ref{fig4} (c).  The insets at the top and bottom of Fig.~\ref{fig4} (b, c)
	depict the extend and SF skin states, which originate from real and complex eigenenergies, respectively. To understand this energy dependence, we examine wavefunction distribution. Given the symmetry relation $|\mathcal{R}(\eta)\rangle=|\mathcal{L}^{\ast}(-\eta)\rangle$, our analysis focuses solely on the right real-space eigenstates, which can be taken as
	$|\mathcal{R}\rangle=\Sigma_{j=1}^{N}|j\rangle\otimes|\psi_{j}\rangle$ with  $|\psi_{j}\rangle=[\mu_{j},\nu_{j}]^{T}$. $\mu(\nu)_{j}$ denotes the occupation amplitude on sublattice a (b) at the $j$-th unit cell. The bulk eigenequation  $H|\mathcal{R}\rangle=E^{o}|\mathcal{R}\rangle$ subsequently yields the following recurrence relations
	\begin{equation}\label{7}
	\begin{aligned}
	&\bm{P}|\psi_{j+1}\rangle+\bm{Q}|\psi_{j-1}\rangle=E^{o}|\psi_{j}\rangle,\\
	&\bm{P}=\begin{bmatrix}J(1+\eta)&t_{2} \\t_{1} &J(1-\eta)\\\end{bmatrix},\,\,\bm{Q}=\begin{bmatrix}J(1-\eta)&t_{1} \\t_{2} &J(1+\eta)\\\end{bmatrix},\\
	\end{aligned}
	\end{equation}
	which indicates that $|\mathcal{R}^{\prime}\rangle =\Sigma_{j=1}^{N}|j\rangle\otimes|\psi_{j}^{\prime}\rangle$ is also an right eigenstate with the eigenvalue $-E^{o}$, where $|\psi_{j}^{\prime}\rangle=(-)|\psi_{j}\rangle$ for odd (even) $j$. By virtue of the identity $H|\mathcal{R}^{\ast}\rangle=E^{o\ast}|\mathcal{R}^{\ast}\rangle$, the four eigenstates $|\mathcal{R}\rangle$, $|\mathcal{R}^{\prime}\rangle$,$|\mathcal{R}^{\ast}\rangle$ and $|\mathcal{R}^{\prime\ast}\rangle$ (only two eigenstates survive when $E^{o}$ is pure imaginary or pure real) exhibit  identical spatial profile, in contrast to the 
	combined $\mathcal{P}\mathcal{T}$  symmetric configuration presented in Sec.~\ref{sec.4}.
	
	The non-trivial solutions to the aforementioned equation can be achieved via the ansatz $|\psi_{j+1}\rangle=\beta|\psi_{j}\rangle$, provided that the characteristic Laurent polynomial satisfies
	\begin{equation}\label{8}
	f(\beta,E^{o}):=\det [\mathcal{H}(\beta)-E^{o}]=0.
	\end{equation}
	Separate $\mathcal{P}$ symmetry $\sigma_{x}\mathcal{H}(\beta)\sigma_{x}=\mathcal{H}(\beta^{-1})$ ensures that Eq.~(\ref{8}) can always be recast as an algebraic equation in terms of $\mathcal{A}=\beta+\beta^{-1}$, explicitly
	\begin{equation}\label{9}
	f(\beta,E^{o})=(J\mathcal{A}-E^{o})^{2}+[t_{0}^{2}\delta^{2}-J^{2}\eta^{2}-t_{0}^{2}]\mathcal{A}^{2}-4(t_{0}^{2}\delta^{2}-J^{2}\eta^{2}).
	\end{equation}
	Here, the two roots of $\mathcal{A}$  can be expressed as
	
	\begin{equation}\label{10}
	\mathcal{A}_{1,2}=(-JE^{o}\mp\sqrt{\mathcal{F}})/\chi,
	\end{equation}
	with $\mathcal{F}=[t_{0}^{2}(1-\delta^{2})+J^{2}\eta^{2}]E^{o2}+4\chi(J^{2}\eta^{2}-t_{0}^{2}\delta^{2})$
	and $\chi= t_{0}^{2}(1-\delta^{2})-J^{2}(1-\eta^{2})$. This means that the roots of $\beta$ must appear in reciprocal pairs as follows:
	
	\begin{equation}\label{11}
	\beta_{i}^{\pm1}=\frac{1}{2}(\mathcal{A}_{i}\pm\sqrt{\mathcal{A}^{2}_{i}-4}),\,and\,|\beta_{i}|=\sqrt{\Bigg|\frac{\mathcal{A}_{i}+\sqrt{\mathcal{A}^{2}_{i}-4}}{\mathcal{A}_{i}-\sqrt{\mathcal{A}^{2}_{i}-4}} \Bigg|}.
	\end{equation}
	For a given $E^{o}$, the four roots $\{\beta_{1},\beta_{1}^{-1},\beta_{2},\beta_{2}^{-1}\}$ are ordered by their moduli such that $|\beta_{1}^{-1}|>|\beta_{2}^{-1}|\geq1\geq|\beta_{2}|>|\beta_{1}|$. In the limit of large $N$, the continuum band condition 
	$\lim _{N\rightarrow\infty}|\beta_{2}|=\lim _{N\rightarrow\infty}|\beta_{2}^{-1}|$ predicted by the non-Bloch band theory \cite{yokomizo2019non, yokomizo2021scaling} remains valid, thus GBZ=BZ. Accordingly, the OBC energy spectra converge to $E^{p}_{\pm}(k)$ as $N$ increases. This feature typically exists in systems with $\mathcal{P}$ symmetry due to the symmetry constraint $f(\beta,E^{o})=f(\beta^{-1},E^{o})$ \cite{yi2020non}. Eq.~(\ref{11}) also implies that the real $\mathcal{A}_{i}$ within the interval $[-2,2]$ uniquely corresponds to $|\beta_{i}|=1$ \cite{lee2020unraveling}, rendering $\mathcal{A}_{i}$ and $|\beta_{i}|$ equivalent for determining eigenstate properties. In this sense, even for a finite open chain, as long as $E^{o}$ is in the continuum energy bands, i.e., $E^{o}\in\{E^{p}_{\pm}(k)\}$, $\beta_{2}$ will reside precisely on the unit circle.

	Since $\lim _{N\rightarrow\infty}|\beta_{2}|=1$ invariably holds for all OPC eigenstates protected by $\mathcal{P}$ symmetry,  the
	traditional exponential skin mode is precluded. Nevertheless, $\lim _{N\rightarrow\infty}|\beta_{2}|=1$ does not necessarily entail $\lim _{N\rightarrow\infty}|\beta_{2}|^{N}=1$. Given that the spatial profile of the SF skin mode remains invariant with respect to system size
	variations, the inequality 
	\begin{equation}\label{12}
	\lim _{N\rightarrow\infty}|\beta_{2}|^{N}<1
	\end{equation}
	can serve as a definitive criterion for its emergence.
	Fig.~\ref{fig4} (b, c) visually demonstrate that the SF skin mode originates from the complex $E^{o}$ in $\mathcal{P}\mathcal{T}$ BP. To substantiate this, we take the eigenstates in the form of a linear combination
	\begin{equation}\label{13}
	\begin{bmatrix}\mu_{j}\\\nu_{j} \\\end{bmatrix}=\beta_{1}^{j}\begin{bmatrix}\mu^{(+1)}\\\nu^{(+1)} \\\end{bmatrix}+\beta_{1}^{-j}\begin{bmatrix}\mu^{(-1)}\\\nu^{(-1)} \\\end{bmatrix}+\beta_{2}^{j}\begin{bmatrix}\mu^{(+2)}\\\nu^{(+2)} \\\end{bmatrix}+\beta_{2}^{-j}\begin{bmatrix}\mu^{(-2)}\\\nu^{(-2)} \\\end{bmatrix}.
	\end{equation}
	The intra-unit-cell
	eigenstates need to satisfy the bulk equation
	\begin{equation}\label{14}
	\mathcal{H}(\beta_{i}^{\pm 1})[\mu^{(\pm i)},\nu^{(\pm i)}]^{T}=E^{o}[\mu^{(\pm i)},\nu^{(\pm i)}]^{T},
	\end{equation}
	which in turn determines the ratio $\nu^{(\pm i)}=\mathcal{X}^{(\pm i)}
	\mu^{(\pm i)}$ between $\mu^{(\pm i)}$ and $\nu^{(\pm i)}$, where
	\begin{equation}\label{15}
	\mathcal{X}^{(\pm i)}=\frac{E^{o}-J(\beta_{i}+\beta_{i}^{-1})\mp J\eta(\beta_{i}-\beta_{i}^{-1}) }{t_{0}(\beta_{i}+\beta_{i}^{-1})\pm  t_{0}\delta(\beta_{i}-\beta_{i}^{-1})},
	\end{equation}
	with the identities
	\begin{equation}\label{16}
	\mathcal{X}^{(+1)}\mathcal{X}^{(-1)}=\mathcal{X}^{(+2)}\mathcal{X}^{(-2)}=1.
	\end{equation} 
	Explicitly, $\mathcal{X}(E^{o})^{(\pm i)\ast}=\mathcal{X}(E^{o\ast})^{(\mp i)}$ if $|\beta_{i}|=1$. While $\mathcal{T}$ symmetry permits real amplitudes for eigenstates,  $\mathcal{P}$ symmetry imposes the constraint $\mu_{j}=\pm\nu_{N+1-j}$ (where the sign arises from $\mathcal{P}^{2}=1$) for any $j$. This ensures all eigenstates possess definite parity, yielding
	\begin{equation}\label{17}
	\nu^{(\pm i)}=\pm\beta_{i}^{\mp(N+1)}\mu^{(\mp i)}.
	\end{equation}
	These eigenstates for a finite chain can be further represented as a superposition of two “plane waves”, $\beta_{1}$ and $\beta_{2}$:
	
	\begin{equation}\label{18}
	\begin{aligned}
	&|\psi_{j}\rangle=|\psi^{1}_{j}\rangle+|\psi^{2}_{j}\rangle,\\
	&|\psi^{1}_{j}\rangle=\mu^{(+1)}\begin{bmatrix}\psi^{1}_{j+}\\\psi^{1}_{j-}\\\end{bmatrix}=\mu^{(+1)}
	\begin{bmatrix}\beta_{1}^{j}\pm\beta_{1}^{N+1-j}\mathcal{X}^{(+1)}\\\beta_{1}^{j}\mathcal{X}^{(+1)}\pm\beta_{1}^{N+1-j}\\\end{bmatrix},\\
	&|\psi^{2}_{j}\rangle=\mu^{(+2)}\begin{bmatrix}\psi^{2}_{j+}\\\psi^{2}_{j-}\\\end{bmatrix}=\mu^{(+2)}\begin{bmatrix}\beta_{2}^{j}\pm\beta_{2}^{N+1-j}\mathcal{X}^{(+2)}\\\beta_{2}^{j}\mathcal{X}^{(+2)}\pm\beta_{2}^{N+1-j}\\\end{bmatrix}.\\
	\end{aligned}
	\end{equation}
	The ratio $\mu^{(+1)}/\mu^{(+2)}$ is governed by the boundary condition $|\psi_{0}\rangle=0$ (or $|\psi_{N}\rangle=0$, as $\mathcal{P}$ symmetry renders these conditions equivalent), which yields the coupled equations $\bm{K}|\Phi\rangle=0$ for the
	coefficients $|\Phi\rangle\equiv[\mu^{(+1)},\mu^{(+2)}]^{T}$, where
	
	\begin{equation}\label{19}
	\bm{K}=\begin{bmatrix}\psi^{1}_{0+}&\psi^{2}_{0+}\\\psi^{1}_{0-} &\psi^{2}_{0-} \\\end{bmatrix}=\begin{bmatrix}1\pm\beta_{1}^{N+1}\mathcal{X}^{(+1)}&1\pm\beta_{2}^{N+1}\mathcal{X}^{(+2)}\\\mathcal{X}^{(+1)}\pm\beta_{1}^{N+1} &\mathcal{X}^{(+2)}\pm\beta_{2}^{N+1} \\\end{bmatrix}.
	\end{equation}
	Nontrivial solutions for $\{\mu^{(+1)},\mu^{(+2)}\}$ necessitate $\det [\bm{K}]=0$. 
	Ignoring the leading terms of $\beta_{1}$ in $\det [\bm{K}]$, we obtain 
	\begin{equation}\label{20}
	\beta_{2}^{N+1}\simeq \pm\frac{\mathcal{X}^{(+1)}-\mathcal{X}^{(+2)}}{1-\mathcal{X}^{(+1)}\mathcal{X}^{(+2)}} \,(a),\,\frac{\mu^{(+1)}}{\mu^{(+2)}}\simeq \frac{\mathcal{X}^{(+2)}-\mathcal{X}^{(-2)}}{\mathcal{X}^{(-2)}-\mathcal{X}^{(+1)}}\,(b).
	\end{equation}  
From Eq.~(\ref{16}), the identities $(\frac{\mathcal{X}^{(+1)}-\mathcal{X}^{(+2)}}{1-\mathcal{X}^{(+
			1)}\mathcal{X}^{(+2)}})(\frac{\mathcal{X}^{(-1)}-\mathcal{X}^{(-2)}}{1-\mathcal{X}^{(-1)}\mathcal{X}^{(-2)}})^{-1}=(\frac{\mathcal{X}^{(+1)}-\mathcal{X}^{(+2)}}{1-\mathcal{X}^{(+
			1)}\mathcal{X}^{(+2)}})$ $(\frac{\mathcal{X}^{(-1)}-\mathcal{X}^{(+2)}}{1-\mathcal{X}^{(-1)}\mathcal{X}^{(+2)}})=1$  suggests that $\mathcal{X}^{(\pm i)}$ in Eq.~(\ref{15}) must be defined such that $|\beta_{2}|^{N+1}<1$  is satisfied.

	Clearly, the inequality $1\geq|\beta_{2}|>|\beta_{1}|$ will make $|\psi_{j}\rangle\approx|\psi^{2}_{j}\rangle$ in Eq.~(\ref{18}) a good approximation, irrespective of minor edge deviations.  Substituting Eq.~(\ref{20}a) into $\psi^{2}_{j+}$ in Eq.~(\ref{18}) facilitates the separation of periodic and non-periodic components:
	\begin{equation}\label{21}
	|\psi^{2}_{j+}|^{2}=2|\varLambda|\cos [2j\theta_{2}-Arg(\varLambda)]+\varPi(j),
	\end{equation}
	where $\varLambda=\frac{\mathcal{X}^{(+1)}-\mathcal{X}^{(+2)}}{\mathcal{X}^{(-2)}-\mathcal{X}^{(+1)}}$, $\beta_{2}=e^{\tau_{2}+i\theta_{2}}$ with $\tau_{2}$ and $\theta_{2}$ being real, and $\varPi(j)=|\beta_{2}|^{2j}+|\beta_{2}|^{-2j}|\varLambda|^{2}$ characterizes the spatial locality of the distribution. A spatial inversion operation yields $|\psi^{2}_{j-}|^{2}$.

	In $\mathcal{P}\mathcal{T}$ UBP, i.e., $|t_{0}\delta|>|J\eta|$, all eigenenergies $E^{o}$ except the zero-energy mode reside within the real PBC continuum spectrum [see Fig.~\ref{fig4} (a)]. Letting $E^{o}\equiv E^{p}_{\pm}(\theta_{2})$, and considering that $\mathcal{A}_{1}^{2}>4$ ($\mathcal{A}_{2}^{2}<4$) with $\mathcal{A}_{i}$ being strictly real,  $\beta_{1}$ ($\beta_{2}$) is constrained to the real axis (the unit circle), corresponding to $\theta_{1}=0$ or $\pi$ ($\tau_{2}=0$) in the parametrization $\beta_{1(2)}=e^{\tau_{1(2)}+i \theta_{1(2)}}$ \cite{lee2020unraveling}.  From Eq.~(\ref{15}),  $ \mathcal{X}^{(+2)}$ can be expressed as 
	\begin{equation}\label{22}
	\mathcal{X}^{(+2)}=\frac{\pm\sqrt{h_{x}^{2}(\theta_{2})+h_{y}^{2}(\theta_{2})-h_{z}^{2}(\theta_{2})}-ih_{z}(\theta_{2})}{h_{x}(\theta_{2})-i h_{y}(\theta_{2})}.
	\end{equation}
	Consequently, $\mathcal{X}^{(+2)}\mathcal{X}^{(+2)\ast}=1$, while the real $\beta_{1}$ ensures that  $\mathcal{X}^{(\pm1)}$ are also real.
	This implies that the wavefunction distribution on both legs remains untilted, as $\varPi(j)$ in Eq.~(\ref{21}) is independent of $j$. Distinct from SF skin modes, the relevant eigenstates $|\mathcal{R}\rangle\approx \sum_{j=1}^{N}|j\rangle\otimes|\psi_{j}^{2}\rangle$ are extended  even for the limited $N$. More generally, in any system possessing  $\mathcal{P}$ symmetry, if a real root $\mathcal{A}_{i}$ of the characteristic polynomial [e.g., Eq.~(\ref{9})] satisfies $\mathcal{A}_{i}^{2}<4$, the corresponding eigenstate is delocalized. 
	
	In $\mathcal{P}\mathcal{T}$ BP, the trajectories of $\beta_{1}^{\pm1}$ form two additional closed loops beyond the BZ. While 
	the OBC eigenstate with an eigenenergy $E^{o}$ within the real PBC spectrum remains extended, for other
	$E^{o}$ that will eventually converge to complex $E^{p}_{\pm}(k)$ [Fig.~\ref{fig4} (b)], $\mathcal{X}^{(+1)}$ ($ \mathcal{X}^{(+2)}$) is generally no longer real (on the unit circle). Based on Eq.~(\ref{15}) and Eq.~(\ref{20}a), we obtain $\lim _{N\rightarrow\infty}|\beta_{2}|^{N}=|\frac{\mathcal{X}^{(+1)}-\mathcal{X}^{(+2)}}{1-\mathcal{X}^{(+1)}\mathcal{X}^{(+2)}}|<1$,  signaling the emergence of an SF skin mode. In this
	regime, the inverse decay length $\tau_{2}$ in $\beta_{2}$ vanishes as $N^{-1}$; however, the skin mode migration from one end to the other, defined as $\mathcal{M}\equiv- N\ln |\beta_{2}|$ converges to a non-zero value, causing the distribution on the two legs to tilt in opposite directions. This scale-independent skewness persists even when the complex $E^{o}$ is far away from PBC spectrum $E^{p}_{\pm}(k)$ at a finite system scale [insets below in Fig.~\ref{fig4} (b)], and is reflected in rescaled profiles  $\varPi(l\equiv\frac{j}{N})=e^{-2\mathcal{M}l}+ e^{2\mathcal{M}l}|\varLambda|^{2}$ in Eq.~(\ref{21}). For sufficiently large systems,  Eq.~(\ref{20}a) suggests that
	\begin{equation}\label{23}
	\mathcal{M}\simeq \frac{N}{N+1} \ln \Big|\frac{1-\mathcal{X}^{(+1)}\mathcal{X}^{(+2)}}{\mathcal{X}^{(+1)}-\mathcal{X}^{(+2)}}\Big|.
	\end{equation}
The
quantity $\mathcal{M}[E^{p}_{\pm}]$ ($=\lim_{N\rightarrow\infty}-N\ln |\beta_{2}|$) represents an extreme migration. In the limit $J\eta\gg1$, we have $\beta_{1}\simeq\frac{(1-\eta)}{(1+\eta)}e^{ik}$ and $\beta_{2}\simeq e^{-ik}$ for large $N$, leading to 
	\begin{equation}\label{24}
	\mathcal{M}[E^{p}_{\pm}]=\ln\Big|\frac{(e^{2ik}-1)[e^{2ik}(1-\eta)^{2}-(1+\eta)^{2}]J\eta}{[e^{4ik}(1-\eta)^{2}-(1+\eta)^{2}]t_{0}} \Big|+\mathcal{O}(\frac{1}{J\eta}),
	\end{equation}
	which can be arbitrarily large, as exemplified by $\mathcal{M}[E^{p}_{\pm}(\pm\pi/2)]\simeq\ln|J(1+\eta^{2})/t_{0}|$.
	To gain an intuitive understanding, we track two OBC eigenenergies in the left half of Fig.~\ref{fig4} (c): One with smallest imaginary part and the other with the smallest real value. These converge to $E^{p}_{-}(k=\frac{\pi}{2})=-1.79 i$ and  $E^{p}_{-}(k=\pi)=-4$, respectively. As illustrated in Fig.~\ref{fig5} (a), the extreme migration $\mathcal{M}[E^{p}_{-}(\pi/2)]=0.867$ ($\mathcal{M}[E^{p}_{-}(\pi)]=0$) corresponds to a (non) tilted state.

		\begin{figure}\label{fig5}
		\scalebox{0.15}[0.15]{\includegraphics{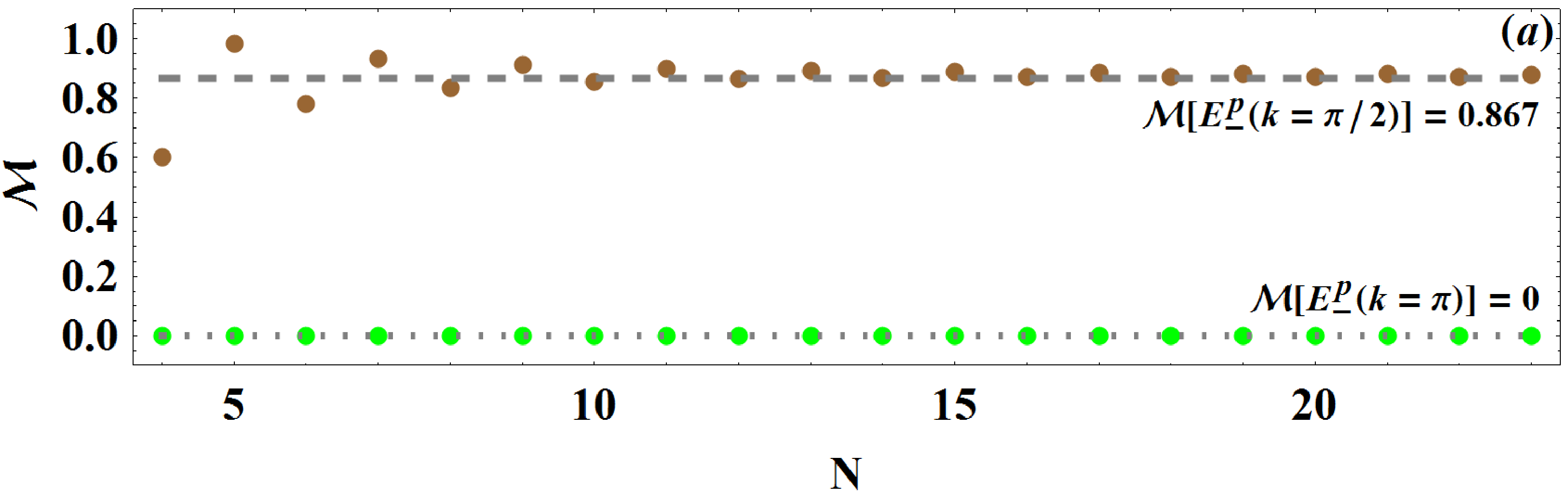}}	
		\scalebox{0.9}[0.9]{\includegraphics{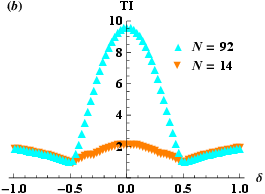}}	\scalebox{0.9}[0.9]{\includegraphics{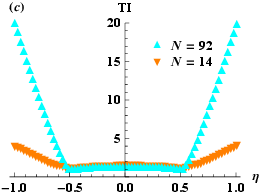}}
		\caption{\label{fig5}(color on line) (a) Migration $\mathcal{M}$ versus $N$: the brown (green) dots track the eigenenergy converging to $E^{p}_{-}(k=\frac{\pi}{2})=-1.79 i$ [ $E^{p}_{-}(k=\pi)=-4$] and the dashed (dotted) line marks the extreme value $\mathcal{M}(E^{p}_{-}(k=\frac{\pi}{2}))$[$\mathcal{M}(E^{p}_{-}(k=\pi)$]. The parameters adopted in (a) are consistent with Fig.~\ref{fig4} (c). (b, c) TI  capturing the $\mathcal{P}\mathcal{T}$ phase transition point $|J\eta|=|t_{0}\delta|$ with
			$\eta=0.5$ (b) and  $\delta=0.5$ (c). $J=t_{0}=1$ is fixed.}
	\end{figure}

	Therefore, we establish that the SF skin modes are uniquely derived from the $\mathcal{P}\mathcal{T}$  symmetry breaking. This $\mathcal{P}\mathcal{T}$-related BBC can be rigorously verified even in systems of moderate size. It is crucial to note, however, that the inverse participation ratio (IPR) fails to identify the SF localization (see App.‌~\ref{a}). Given that  the skin
	modes on two legs are localized at opposite ends, we introduce the total imbalance (TI) \cite{wu2024atomic} to quantify this spatial localization asymmetry:
	\begin{equation}\label{25}
	TI=\sum^{N}_{\ell=1}\dfrac{\sum_{\lambda=\mu,\nu}|\sum_{i\leq N/2}|\lambda_{i}^{\ell}|^{2}-\sum_{i\geq N/2}|\lambda_{i}^{\ell}|^{2}|}{\sum^{N}_{i=1}|\mu_{i}^{\ell}|^{2}+|\nu_{i}^{\ell}|^{2}},
	\end{equation}
	here the superscript ``$\ell$'' denotes $\ell$-th right eigenstate $|\mathcal{R}_{\ell}\rangle$.  Fig.~\ref{fig5} (b) and (c) show that the TI is indeed a reliable indicator even at moderate $N$ [ i.e., $N\sim O(10^{1})$]. Notably, the critical point of the TI, marking the transiton from Bloch-wave-like extended states to SF skin modes, coincides precisely with the $\mathcal{P}\mathcal{T}$ phase transition point of $\mathcal{H}(k)$ defined in Eq.~(\ref{2}).

	\begin{figure}\label{fig6}
		\scalebox{0.124}[0.124]{\includegraphics{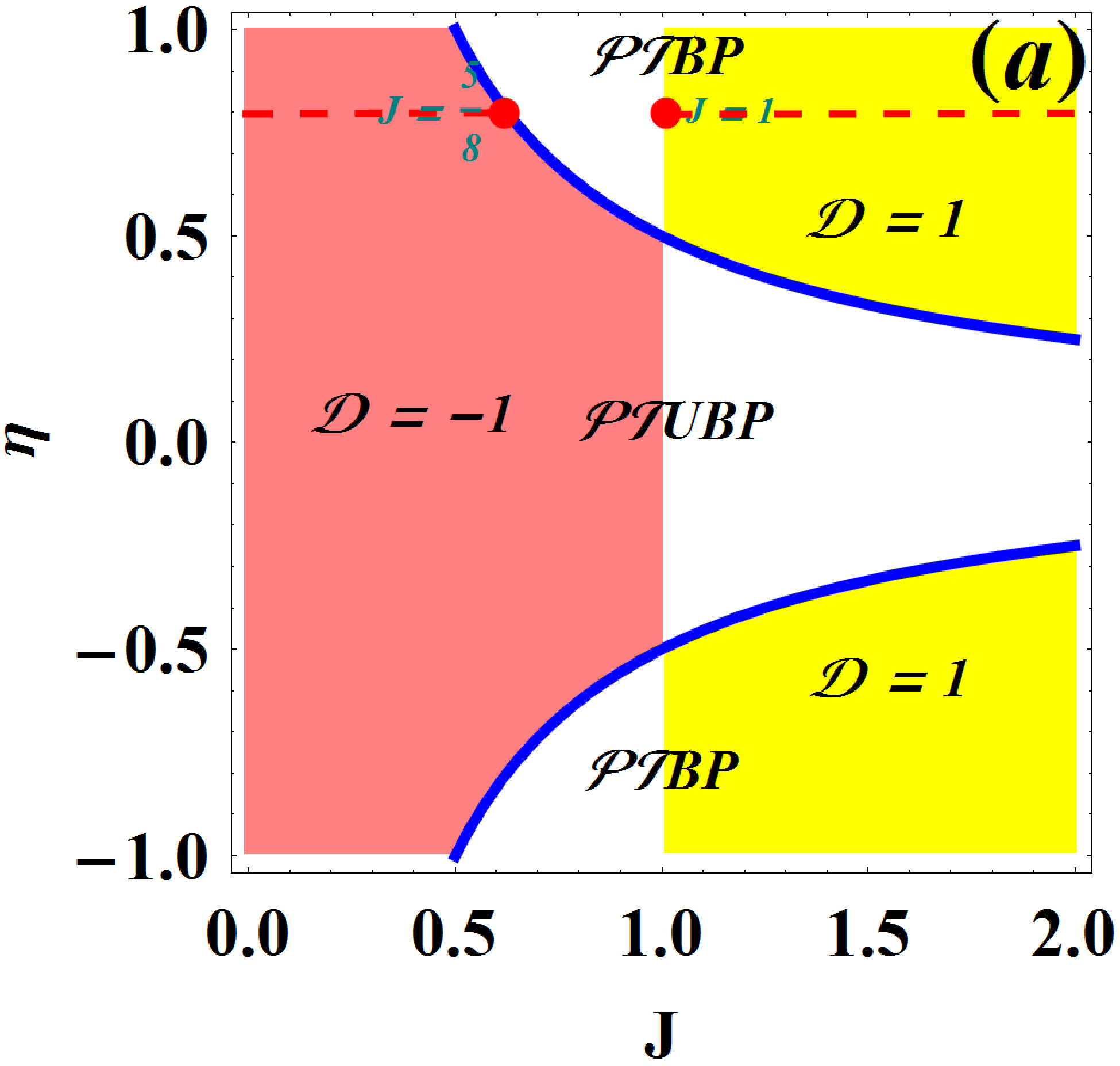}}
		\scalebox{0.14}[0.14]{\includegraphics{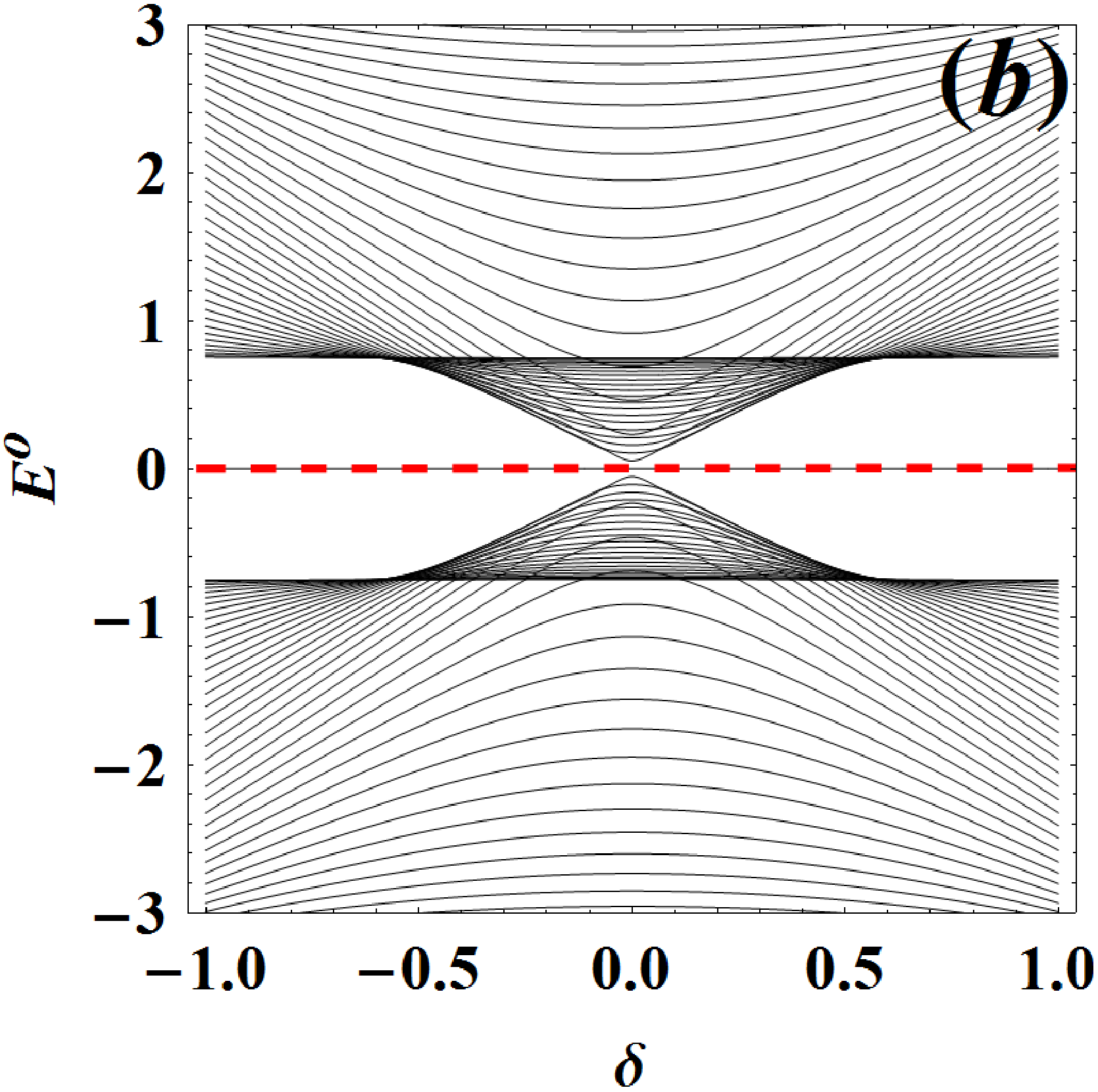}}
		\scalebox{0.1263}[0.1263]{\includegraphics{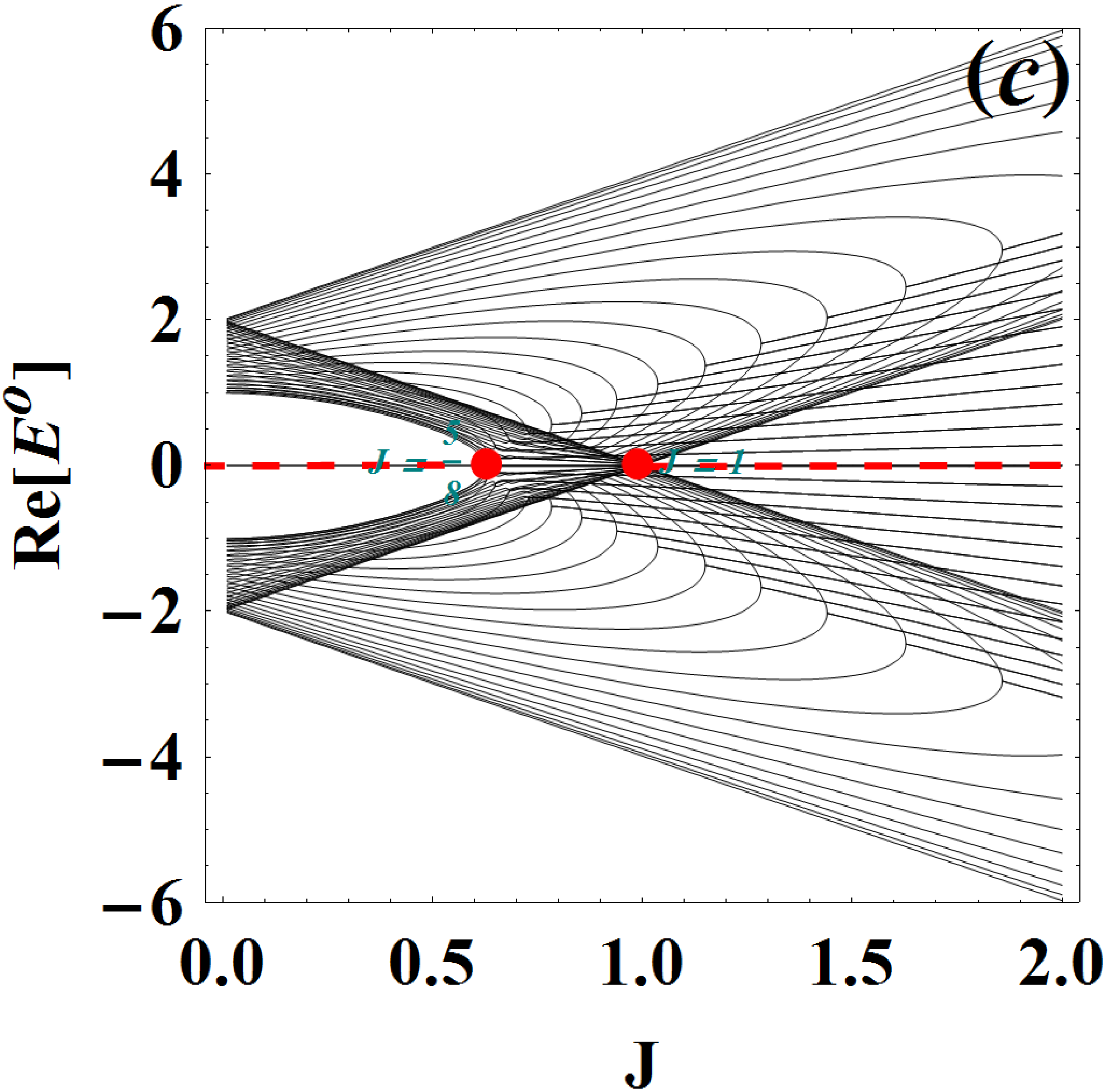}}	
		\scalebox{0.126}[0.126]{\includegraphics{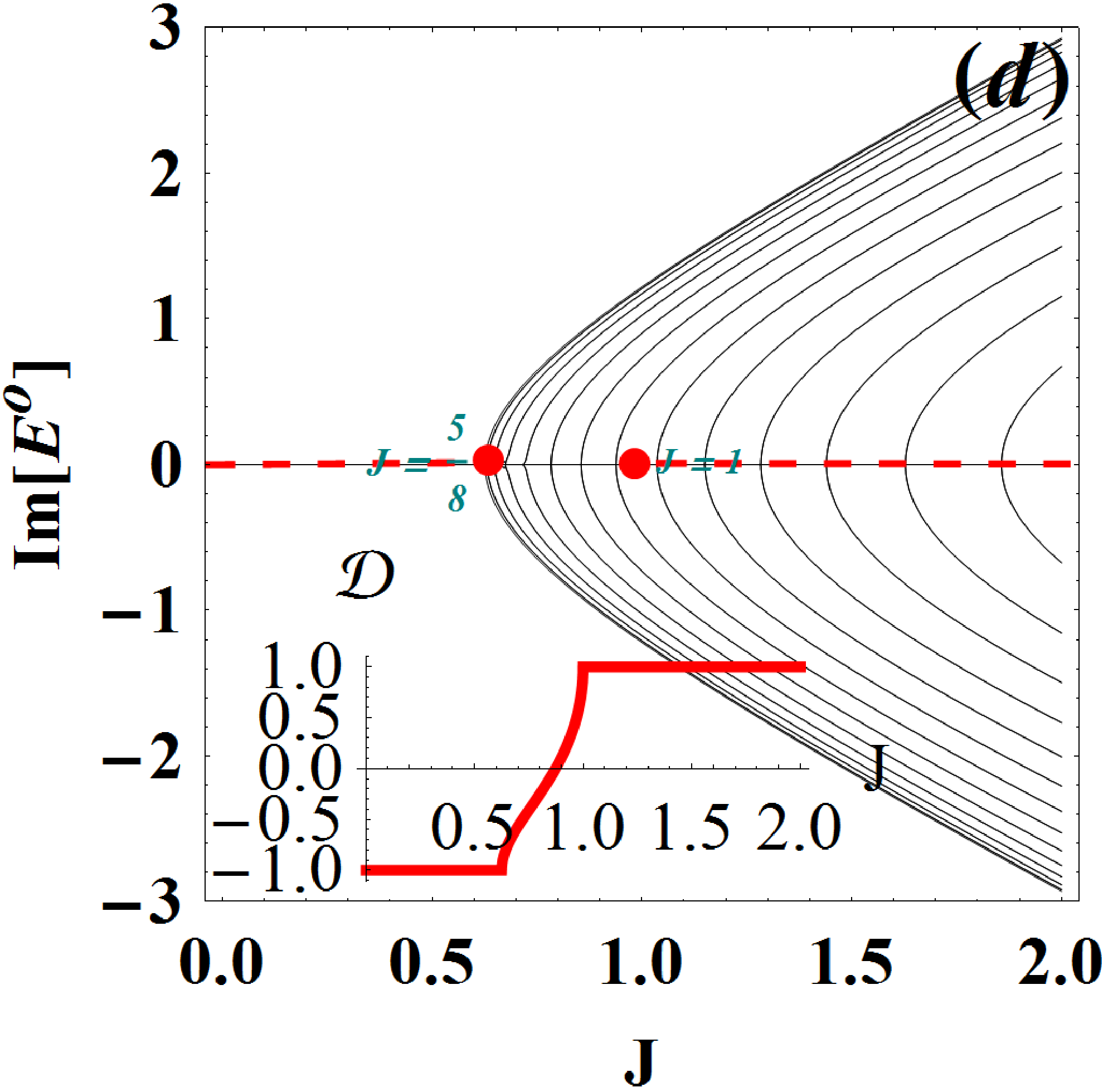}}	
		\caption{\label{fig6}(color on line) (a) Phase diagram of the $\mathcal{P}$-symmetric Non- Hermitian ladder on the $J-\eta$ plane with $t_{0}=1$ and $\delta=0.5$.   The topological phase lies in the yellow ($\mathcal{D}=1$) and pink ($\mathcal{D}=-1$) regions, while the white region indicates the topological trivial phase. The blue line represents the phase boundary of the $\mathcal{P}\mathcal{T}$ phase transition. The endpoints (red dots) of  the zero-mode (red dashed) line ($\eta=0.8$) are two topological transition points ( $J=\frac{5}{8}$ and $1$). (b, c, d) OBC energy spectrum
			in Hermitian [$\eta=0$, $J=\frac{5}{8}$ (b)] and non-Hermitian [$\eta=0.8$ (c, d)] cases with $N=43$. In the inset of (d), $\mathcal{D}$ undergoes a sudden change at topological transition points.}
	\end{figure}

	\subsection{Topological BBC}

	In practice, topologically protected edge modes are expected to remain outside the bulk energy band under OBC in the thermodynamic limit, thus preventing the coalescence of OBC spectrum. In our Hermitian case, this can be achieved by reducing the intensity of the energy offset $h_{0}$, so that the appearance of edge modes bears no relation to winding numbers.  In fact, under the established topological classification of non-Hermitian systems with reflection symmetry, the Hamiltonian in Eq.~(\ref{2}) falls into the $AI(R^{+})$ class, which is ‌in principle‌ supposed to characterized by a $\mathbb{Z}_{2}$ topological invariant $D=sgn\{\det [\mathcal{H}(k)]\}$ ($sgn\{\cdot\}$ picks up the sign of ``$\cdot$'') \cite{liu2019topological}. This invariant generalizes the earlier
	 $\mathbb{Z}_{2}$ quantity proposed for $\mathcal{P}\mathcal{T}$-symmetric zero-dimensional non-Hermitian systems \cite{gong2018topological}. However, prior work \cite{liu2019topological} has not identified any $AI(R^{+})$ class example where  $D$ reliably reflect the existence of topological edge modes.  For Hamiltonians in the form given in Eq.~(\ref{2}),  $h_{0}^{2}(k)+h_{z}^{2}(k)$ [$h_{x}^{2}(k)+h_{y}^{2}(k)$] represents the square of the chain's intrinsic energy (the interchain interaction energy).
	 As long as their difference $\det [\mathcal{H}(k)]=[h_{0}^{2}(k)+h_{z}^{2}(k)]-[h_{x}^{2}(k)+h_{y}^{2}(k)]$ is large enough, a small perturbations will not alter the sign of $\det [\mathcal{H}(k)]$.
	 For greater rigor, we transformed the $D$ into a more robust form $\mathcal{D}=\int^{\pi}_{-\pi}sgn\{\det [\mathcal{H}(k)]\}\frac{dk}{2\pi}$. 
	Substituting Eq.~(\ref{2}) into $\mathcal{D}$ yields:
	\begin{equation}\label{26}
	\begin{aligned}
	\mathcal{D}=
	&\left\{
	\begin{aligned}
	=1,\,\,\,\,\,\,\varDelta>0\,and\,|J\eta|>|t_{0}\delta|,\\
	=-1,\varDelta>0\,and\,|J\eta|<|t_{0}\delta|,\\
	\in(-1,1),\,\,\,\,\,\,\,\,\,\,\,\,\,\,\,\,\,\,\,\,\,\,\,\,\,\,\,\,\,\,\,\,\,\,\,\,\,\,\,\,\,\,\,\,\varDelta<0,
	\end{aligned}
	\right.\\
	\end{aligned}
	\end{equation}
	where $\varDelta=(J^{2}\eta^{2}-t_{0}^{2}\delta^{2})/(J^{2}-t_{0}^{2})$ regardless of the phase boundary $J=t_{0}$. Remarkably, the $\mathbb{Z}_{2}$ invariant $\mathcal{D}$ accurately predicts the zero-energy edge modes because the topological BBC is guaranteed by
	a hidden chiral symmetry, which enforces $\det [\mathcal{H}(k)]=\det [\mathcal{H}(\pi-k)]$ (see App.‌~\ref{b}).
	
	Complex energies belonging to different bands cannot intersect due to separation by
	the real axis. Overlapping real energies $E^{p}_{-}(k_{-})=E^{p}_{+}(k_{+})$ for arbitrary $k_{\pm}$ requires:
	\begin{equation}\label{27}
	\begin{aligned}
	&\left\{
	\begin{aligned}
	[E^{p}_{-}(0),E^{p}_{-}(k_{1})]\cap[E^{p}_{+}(k_{2}),E^{p}_{+}(\pi)]\neq\varnothing,\,|J\eta|>|t_{0}\delta|,\\
	[E^{p}_{-}(\pi),E^{p}_{-}(0)]\cap[E^{p}_{+}(\pi),E^{p}_{+}(0)]\neq\varnothing,\,\,\,\,\,\,\,\,|J\eta|<|t_{0}\delta|,
	\end{aligned}
	\right.\\
	\end{aligned}
	\end{equation}
where $k_{1,2}$ are two EPs as dipicted in Fig.~\ref{fig3} (b). This leads to $t_{0}>J$ ($t_{0}<J$) in $\mathcal{P}\mathcal{T}$BP (UBP) [white region in  Fig.~\ref{fig6} (a)]. Conversely, band isolation occurs exclusively in the $\mathcal{D}=\pm 1$ region, where edge states reside within the bandgap. While the overlap of two PBC bands does not guarantee a genuine band degeneracy, i.e. $E^{p}_{-}(k)=E^{p}_{+}(k)$, this distinction is completely  redundant for OBC spectra. Fig.~\ref{fig6} shows that $\mathbb{Z}_{2}$ topological phases exist in both the $\mathcal{P}\mathcal{T}$ broken and unbroken regions. A pair of zero modes  appears at the boundary $|J\eta|=|t_{0}\delta|$ only when $N$ is odd. More uniquely, 
	at the triple points $|\delta|=|\eta|$ and $J=t_{0}$ (also a OBC EP for odd $N$), $\det [\mathcal{H}(k)]=0$ and two dispersive bands merge to form an entire flat band: $E^{p}_{-}[k\in\pm(\frac{\pi}{2},\pi)]=E^{p}_{+}[k\in\pm(0,\frac{\pi}{2})]=0$. This flat band is singular due to the band crossing \cite{rhim2019classification}, and accordingly, the matrices $\bm{P}$ and $\bm{Q}$ in Eq.~(\ref{7}) are no longer of full rank.
	This macroscopic degeneracy supports two types of compact localized zore-modes with different occupied sizes ($U=1\,or\,2$) \cite{flach2014detangling,maimaiti2019universal}:
	\begin{equation}\label{28}
	\begin{aligned}
	&|\mathcal{R}_{j}(\delta=\eta)\rangle=|j\rangle\otimes[\frac{1}{\sqrt{2}},\frac{-1}{\sqrt{2}}]^{T},\,j=1,2\ldots N,\\
	&|\mathcal{R}_{1}(\delta=-\eta)\rangle=|1\rangle\otimes|\phi_{2}\rangle,\,\,\,\,|\mathcal{R}_{N}(\delta=-\eta)\rangle=|N\rangle\otimes|\phi_{1}\rangle,\\
	&|\mathcal{R}_{j}(\delta=-\eta)\rangle=\frac{|j-1\rangle\otimes|\phi_{1}\rangle-|j+1\rangle\otimes|\phi_{2}\rangle}{\sqrt{2}},\,j=2,3\ldots N-1,\\
	\end{aligned}
	\end{equation}
	where $|\phi_{1}\rangle=[\frac{\eta-1}{\sqrt{2(1+\eta^{2})}},\frac{\eta+1}{\sqrt{2(1+\eta^{2})}}]^{T}$ and $|\phi_{2}\rangle=\sigma_{x}|\phi_{1}\rangle$.
	Unlike other localization
	mechanisms featuring decaying tails, the modes $|\mathcal{R}_{j}(\delta=\pm\eta)\rangle$ in Eq.~(\ref{28}) exhibit highly confined spatial support and are strictly localized on selected lattice sites. While nonreciprocity here is not necessary for flat band formation, it modulates the amplitude distribution between sublattices in $|\mathcal{R}_{j}(\delta=-\eta)\rangle$, thereby facilitating the manipulation of localization.

To reveal the underlying mechanism of the topological BBC, we revisit the recursion formula Eq.~(\ref{7}) with  $E^{o}=0$ imposed, which reduces to
	\begin{equation}\label{29}
	|\psi_{j+1}\rangle=\bm{T}|\psi_{j-1}\rangle,
	\end{equation}
	where the transfer matrix \begin{equation}\label{30}
	\bm{T}=-\bm{P}^{-1}\bm{Q}=
	\begin{bmatrix}\frac{t_{2}^{2}-J^{2}(1-\eta)^{2}}{J(1-\eta^{2})-t_{1}t_{2}}&\frac{J(1+\eta)t_{2}-J(1-\eta)t_{1}}{J^{2}(1-\eta^{2})-t_{1}t_{2}} \\\frac{J(1-\eta)t_{1}-J(1+\eta)t_{2}}{J^{2}(1-\eta^{2})-t_{1}t_{2}} &\frac{t_{1}^{2}-J^{2}(1+\eta)^{2}}{J^{2}(1-\eta^{2})-t_{1}t_{2}}\\\end{bmatrix}
	\end{equation}
belongs to $SL(2,\mathbb{R})$ group, and $\bm{P}^{-1}$ denots the inverse matrix of $\bm{P}$. The properties of the zero-mode solution are embedded in the eigenvalue equation $\bm{T}|\psi_{\pm}\rangle=\mathcal{Z}_{\pm}|\psi_{\pm}\rangle$, where the two eigenstates take the forms:
	\begin{equation}\label{31}
	|\psi_{\pm}\rangle=[\mu_{\pm},-1
	]^{T}
	\end{equation}
	with
	\begin{equation}\label{32}
	\mu_{\pm}=\frac{t_{0}^{2}\delta+J^{2}\eta\mp\sqrt{(J^{2}-t_{0}^{2})(J^{2}\eta^{2}-t_{0}^{2}\delta^{2})}}{J t_{0}(\delta+\eta)},
	\end{equation}
	and $\mu_{+}\mu_{-}=1$ embodies the $\mathcal{P}$ symmetry. The two eigenvalues are given by
	
	\begin{equation}\label{33}
	\mathcal{Z}_{\pm}=\dfrac{\sqrt{\varDelta}\pm 1}{\sqrt{\varDelta}\mp 1}\,\,\,\,\,for\,J>t_{0},\,\,\mathcal{Z}_{\pm}=\dfrac{\sqrt{\varDelta}\mp 1}{\sqrt{\varDelta}\pm 1}\,\,\,\,\,for\,J<t_{0},
	\end{equation}
	which determine the localization lengths of the two
	edge states and satisfy $\mathcal{Z}_{+}\mathcal{Z}_{-}=1$. Whenever 
	$\varDelta<(>)0$ implying $|\mathcal{Z}_{\pm}|=(\neq)1$, the two zero-energy modes are extended (localized). The condition $|\mathcal{Z}_{\pm}|=0$ or $\infty$ at $\varDelta=1$ enables the extreme localization.
	For any odd $N$, $|\psi_{j}\rangle=0$ (here $j=0,2,\cdots N+1$) fully enforces the specified boundary conditions,  yielding the exact analytical solutions for a pair of zero-energy modes:
	\begin{equation}\label{34}
	|\mathcal{R}_{\pm}\rangle=\frac{1}{\mathcal{C_{\pm}}}\sum_{j=1}^{N}|j\rangle\otimes\mathcal{Z}_{\pm}^{\frac{j-1}{2}}\Lambda_{-}(j)|\psi_{\pm}\rangle,
	\end{equation}
		
	where $C_{\pm}$ ($C_{-}=\mathcal{Z}_{-}^{\frac{N-1}{2}}C_{+}$) are normalization constants, and $\Lambda_{\pm}(j)=(\frac{1\pm e^{ij\pi}}{2})$ accounts for the staggered occupation. In the topologically protected region $\mathcal{D}=\pm1$, these modes are
	exponentially squeezed towards the boundaries.
	For even $N$, no
	zero mode presents in $\mathcal{D}\neq\pm1$ region; however, the
	$\mathbb{Z}_{2}$ invariant $\mathcal{D}=\pm1$ supports a pair of  edge modes with the eigenvalues rapidly approaching zero as $N$ increases, well approximated by:
	
\begin{equation}\label{35}
	\begin{aligned}
	|\mathcal{R}^{\prime}_{\pm}\rangle\simeq
	&\left\{
	\begin{aligned}
	\frac{1}{\mathcal{C}^{\prime}_{\pm}}\sum_{j=1}^{N}|j\rangle\otimes\mathcal{Z}_{\pm}^{\frac{2j-3 \mp 1}{4}}\Lambda_{\pm}(j)|\psi_{\pm}\rangle,\,\,\,\,\,\,|\mathcal{Z}_{-}|<1,\\
	\frac{1}{\mathcal{C}^{\prime}_{\pm}}\sum_{j=1}^{N}|j\rangle\otimes\mathcal{Z}_{\pm}^{\frac{2j-3 \pm 1}{4}}\Lambda_{\mp}(j)|\psi_{\pm}\rangle,\,\,\,\,\,\,|\mathcal{Z}_{-}|>1,
	\end{aligned}
	\right.\\
	\end{aligned}
	\end{equation}
	even for small $N\approx \mathcal{O}(10^{1})$, where   $\mathcal{C}^{\prime}_{-}=\mathcal{Z}_{-}^{\frac{N-2}{2}}\mathcal{C}^{\prime}_{+}$.  $\mathcal{P}$ symmetry dictates that the edge  modes appear in pairs, ensuring that 
	both $|\mathcal{R}_{\pm}\rangle$ and $|\mathcal{R}^{\prime}_{\pm}\rangle$ possess well-defined parity: $|\mathcal{R}_{+}^{(\prime)}\rangle=\mathcal{P}|\mathcal{R}_{-}^{(\prime)}\rangle$. Ultimately, our model successfully fills the gap in the topological BBC of the $AI(R^{+})$ class.

	\section{ combined $\mathcal{P}\mathcal{T}$ symmetry}\label{sec.4}

We now reintroduce balanced gain-loss while imposing the constraint $\eta_{a}=-\eta_{b}=\eta$, which dictates that the term $h_{z}$ in Eq.~(\ref{2}) is updated as follows:
\begin{equation}\label{36}
	h_{z}=2J\eta\sin(k)
	\longrightarrow 
	h_{z}=2J\eta\sin(k)+\gamma.
\end{equation}
In this scenario, individual $\mathcal{P}$ and $\mathcal{T}$ symmetries are broken, but the combined $\mathcal{P}\mathcal{T}$ symmetry remains intact. We will see that breaking the $\mathcal{P}\mathcal{T}$ symmetry induces the exponential skin mode rather than the SF skin mode, with a $\gamma$-modulated localization direction. Since the hidden chiral symmetry persists, a modified topological $\mathbb{Z}_{2}$ invariant remains valid.	
	
\begin{figure}[H]\label{fig7}
	\scalebox{0.256}[0.256]{\includegraphics{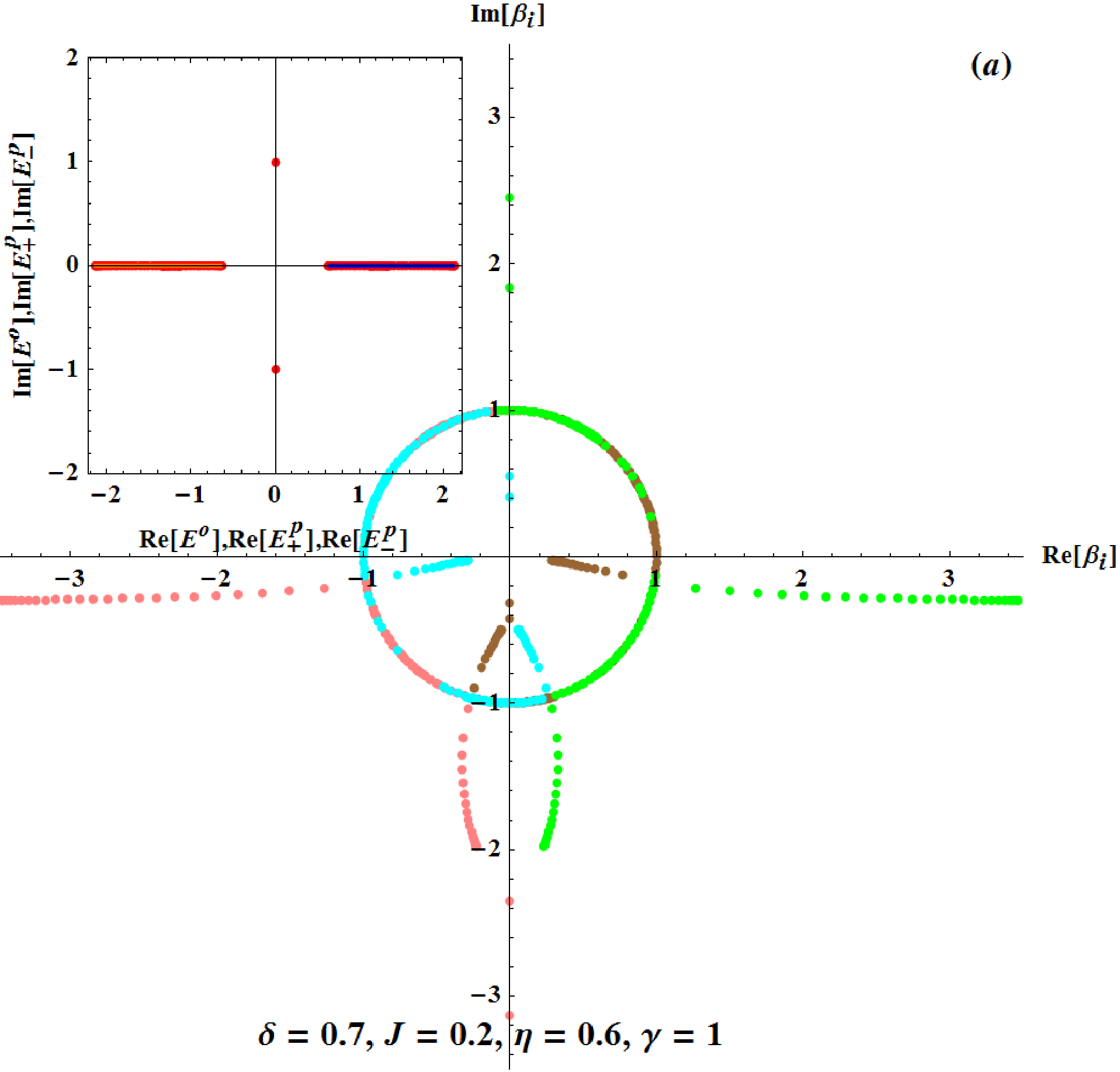}}
	\scalebox{0.24}[0.24]{\includegraphics{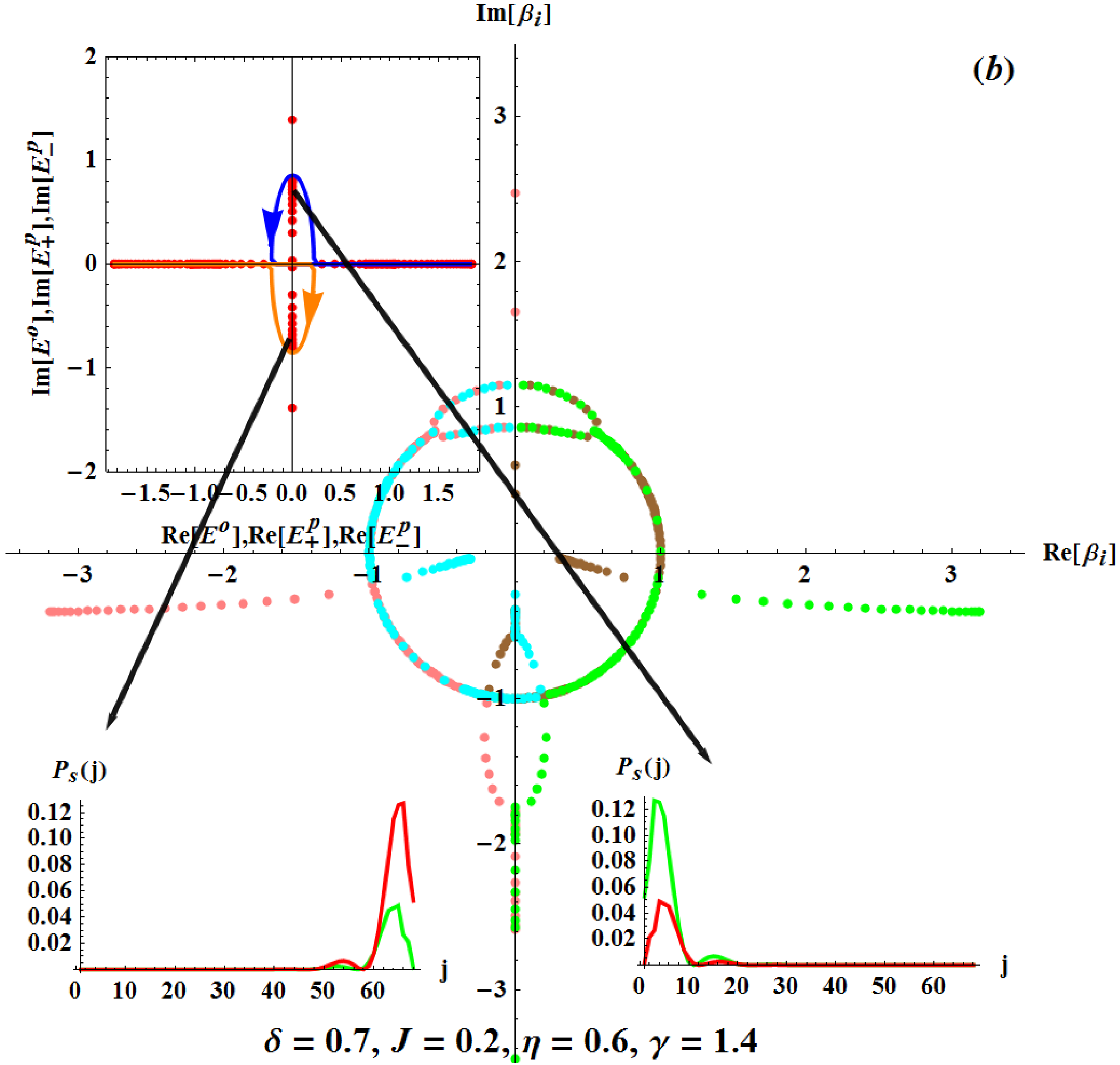}}
	
	\caption{\label{fig7}(color on line) Numerical GBZ plus energy spectra (upper left insets) in $\mathcal{P}\mathcal{T}$ UPB (a) and in $\mathcal{P}\mathcal{T}$ PB (b) at $N=68$.  $\beta_{1}$(cyan dots), $\beta_{2}$ (green dots), $\beta_{3}$ (pink dots), $\beta_{4}$ (brown dots), $E^{p}_{-}(k)$ (orange line), $E^{p}_{+}(k)$ (blue line), $E^{o}$ (red dots). The purely imaginary  OBC eigenenergies $E^{o}_{+}$ ($E^{o}_{-}$) lying on the positive (negative) imaginary axis are circled counterclockwise (clockwise) by $E^{p}_{+}(k)$ [$E^{p}_{-}(k)$].  The orange or blue arrowheads indicate the direction of increasing $k$ as shown in the upper-left inset of (b).
		$E^{o}_{+}\approx0.72i$	corresponds to $|\beta_{1+}|=|\beta_{2+}|\approx0.88$, $|\beta_{3+}|\approx2.45$, and $|\beta_{4+}|\approx0.53$, pointing towards a left-localized skin mode (lower-right inset). Its partner is a right-localized skin mode  (lower-left inset) with  $E^{o}_{-}\approx-0.72i$ and $|\beta_{j-}|\approx|\beta_{j+}|^{-1}$.}
\end{figure}

	\subsection{ $\mathcal{P}\mathcal{T}$-related and non-Hermitian BBCs}
	The AWN, regardless of the specific form of $h_{i}$ in Eq.~(\ref{2}), reliably  predicts  the $\mathcal{P}\mathcal{T}$ phase transition. Gain-loss alone cannot generate any skin modes due to the particle-hole symmetry breaking \cite{yi2020non}, so we maintain the nonreciprocity based on the parameters in Fig.~\ref{fig4} (a), and then strengthen $\gamma$ gradually. For fixed $\delta=0.7$, $J=0.2$, and $\eta=0.6$ within the $\varGamma=-1$  regime of Fig.~\ref{fig2} (a),  the $\mathcal{P}\mathcal{T}$ phase transition point is at $\gamma^{\ast}\approx\pm1.16$. Prior to the  $\mathcal{P}\mathcal{T}$ transition, the GBZ develops new branches while preserving the  unit circle, and the bulk PBC and OBC spectra completely overlap [inset of Fig.~\ref{fig7} (a)], indicating all eigenstates extend over the bulk. A larger $|\gamma|$  breaks $\mathcal{P}\mathcal{T}$ symmetry ($\varGamma>-1$), driving a clear discrepancy between OBC and PBC spectra, and  rendering the GBZ unit circle incomplete [see Fig.~\ref{fig7} (b)], such that partial bulk modes associated with the arc off the unit circle become exponentially localized at the boundary. 
	If the $\mathcal{P}\mathcal{T}$ symmetry is already broken at $\gamma=0$ [Fig.~\ref{fig4} (b, c)], gain-loss will further convert SF skin modes into exponential ones.  Upon complete $\mathcal{P}\mathcal{T}$ symmetry breaking, a sufficiently large $|\gamma|$ opens a line gap between bands, causing a radical departure of the GBZ from the unit circle and full localization of the bulk modes. The $\mathcal{P}\mathcal{T}$ symmetry ensures $
\det [\mathcal{H}(\beta)-E^{o}]=\det [\mathcal{H}(\beta^{-1})-E^{o\ast}]=0$,  leading to a pair of skin modes with complex conjugate eigenenergies localized at opposite ends. These are correctly predicted by the spectral winding numbers of the two PBC bands:
\begin{equation}\label{37}
\mathcal{W}^{\pm}_{s}=
\int^{\pi}_{-\pi} \partial_{k}Arg[E^{p}_{\pm}(k)-E^{o}_{\pm}]=\pm sgn[\gamma],
	\end{equation}
where $E^{o}_{\pm}$  denotes the complex energy enclosed  by $E^{p}_{\pm}(k)$ [upper-left inset of Fig.~\ref{fig7} (b)].

	\subsection{Robust topological BBC protected by hidden chiral symmetry}
	
With the inclusion of $\gamma$, the recursion equation Eq.~(\ref{7}) becomes:
	\begin{equation}\label{38}
	\bm{P}|\psi_{j+1}\rangle+\bm{Q}|\psi_{j-1}\rangle=(E^{o}-i\gamma\sigma_{z})|\psi_{j}\rangle.
	\end{equation}
Here, $\gamma$ shifts the zero modes towards the imaginary axis: $E^{o}=0\rightarrow E^{o}=\pm i\gamma$ (isolated red dots in upper-left insets of Fig.~\ref{fig7}). As $|\gamma|$ increases, these two isolated energy points merge into the OBC spectra. Since the boundary localization still depends on the transfer matrix $\bm{T}=-\bm{P}^{-1}\bm{Q}$, we can artificially remove the gain-loss component from the original   definition  of the $\mathbb{Z}_{2}$ invariant $\mathcal{D}$ to achieve an exclusive BBC for the topological edge modes. 
In this sense,  the emergence of NHSE does not necessitate redefining  topological invariants based on non-Bloch theory.

	\section{absence of $\mathcal{P}\mathcal{T}$ symmetry}\label{sec.5}
	Setting $\eta_{a}\neq-\eta_{b}$ and $\gamma=0$ preserves $\mathcal{T}$ symmetry but breaks  $\mathcal{P}$ symmetry, resulting in the absence of the combined $\mathcal{P}\mathcal{T}$ symmetry. This requires the following substitution in Eq.~(\ref{2}):
	\begin{equation}\label{39}
	\begin{matrix}h_{0}=2J\cos(k),\,\,\,\, \\h_{z}=2J\eta\sin(k),\end{matrix}
	\longrightarrow 
	\begin{matrix}h_{0}=2J\cos(k)+iJ(\eta_{a}+\eta_{b})\sin(k), \\h_{z}=J(\eta_{a}-\eta_{b})\sin(k).\,\,\,\,\,\,\,\,\,\,\,\,\,\,\,\,\,\,\,\,\,\,\,\,\,\,\,\,\,\,\,\,\,\,\,\,\,\,\,\,\,\,\,\,\end{matrix}
	\end{equation}
	In this regime, exponential skin modes are ubiquitous due to the persistent $\mathcal{P}\mathcal{T}$ symmetry breaking, decoupling the AWN's prediction of PBC band degeneracy from the skin effect.
	Further assuming $\eta_{a}=\eta_{b}=\eta$, the pseudo-inversion symmetry $\sigma_{x}\mathcal{H}(k)\sigma_{x}=\mathcal{H}^{\dagger}(-k)$ emerges, reducing the AWN to its normal form. We will subsequently analyze the localized features of bulk modes at $\delta=0$, followed by further elucidation of the topological boundary states.

	\subsection{non-Hermitian BBC and Abnormal skin modes}

	To highlight the abnormal bulk modes, we restore the sublattice symmetry $\sigma_{x}\mathcal{H}(k)\sigma_{x}=\mathcal{H}(k)$ by setting $\delta=0$. Under this condition, the PBC energy bands
	\begin{equation}\label{40}
	E^{p}_{\pm}(k)=2[J\cos (k)+i\eta J\sin (k)\pm t_{0}|\cos (k)|]
	\end{equation}
become degenerate at $k=\pm\frac{\pi}{2}$.
	The four roots of the corresponding characteristic polynomial can be divided into two pairs:
	\begin{equation}\label{41}
	\beta_{1,4}=\frac{E^{o}\pm\sqrt{E^{o2}-4d_{-}}}{2(J-t_{0}+J\eta)},\,\beta_{2,3}=\frac{E^{o}\pm\sqrt{E^{o2}-4d_{+}}}{2(J+t_{0}+J\eta)},
	\end{equation}
	with $d_{\pm}=(J\pm t_{0})^{2}-(J\eta)^{2}$. Each pair of roots satisfies the following product relationship:
	\begin{equation}\label{42}
	\beta_{1}\beta_{4}=\frac{J-t_{0}-J\eta}{J-t_{0}+J\eta},\,\, \beta_{2}\beta_{3}=\frac{J+t_{0}-J\eta}{J+t_{0}+J\eta}.
	\end{equation}
	This simplified Hamiltonian $H(\delta=\gamma=0,\eta_{a}=\eta_{b}=\eta)$ has two sets of OBC spectra (App.‌~\ref{c})

	\begin{figure}\label{fig8}
		\scalebox{0.151}[0.151]{\includegraphics{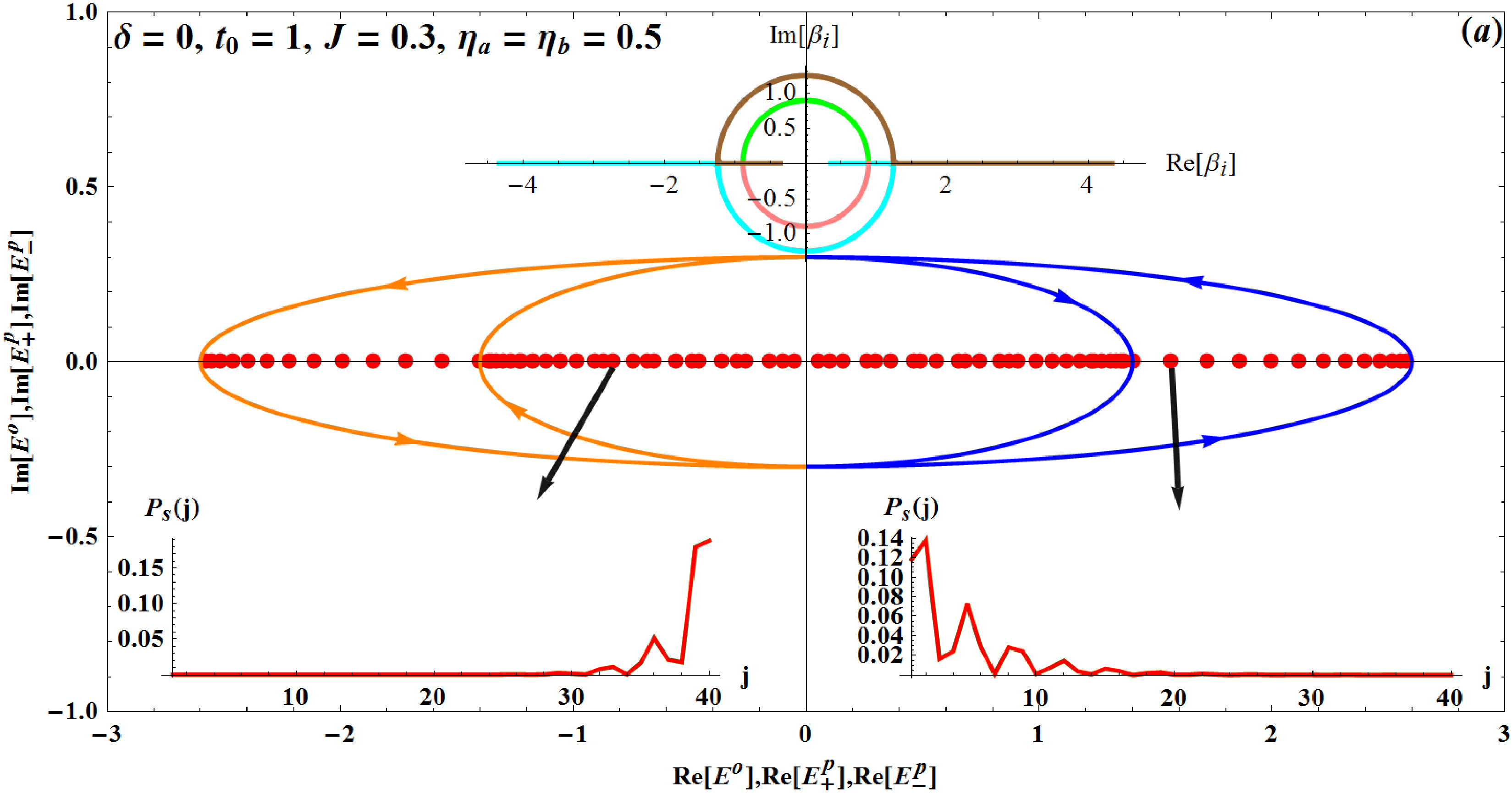}}
		\scalebox{0.149}[0.149]{\includegraphics{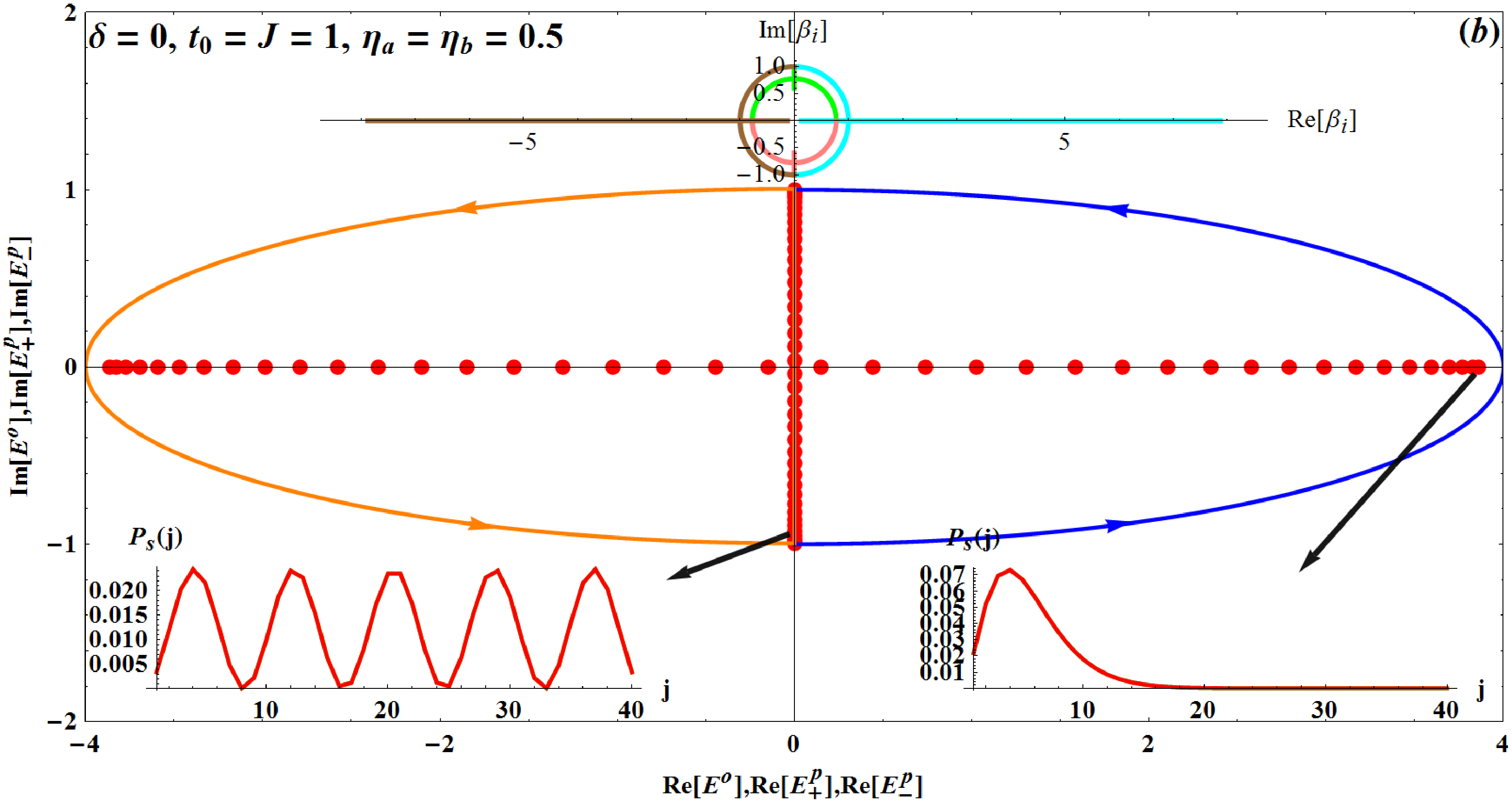}}
		
		\caption{\label{fig8}(color on line) PBC energy spectra $E^{p}_{-}(k)$ (orange line), $E^{p}_{+}(k)$ (blue line) and OBC energy spectra $E^{o}$ (red dots) for $N=40$. Top insets: exact GBZs with $\beta_{1}$ (cyan), $\beta_{2}$ (green), $\beta_{3}$ (pink), $\beta_{4}$ (brown). $r_{1}=\sqrt{\frac{17}{11}}$,
			$r_{2}=\sqrt{\frac{23}{29}}$ (a).  $r_{1}=1$, $r_{2}=\sqrt{\frac{3}{5}}$  (b). Bottom insets: the eigenstate profiles exhibit both normal and anomalous modes at $E^{o}\approx1.57$ (a), $\approx3.86$ (b) and $E^{o}\approx-0.83$ (a), $\approx-0.93 i$ (b), respectively. The arrows on the curve $E^{p}_{\pm}(k)$ indicate the direction of increasing $k$.}
	\end{figure}

	\begin{equation} \label{43}
	E^{o}_{\pm j}=2\sqrt{d_{\pm}}\cos (\frac{j\pi }{N+1}),\,\,\,j=1,2,\cdots N.
	\end{equation}
	Especially when $d_{-}<0$, the energy $E^{o}_{- j}$ is purely imaginary. Substituting Eq.~(\ref{43}) into Eq.~(\ref{41}) produces the GBZ [top insets in Fig.~\ref{fig8} (a, b)] that splits into  two consecutive circles with radii $r_{1}=\sqrt{|\frac{J-t_{0}-J\eta}{J-t_{0}+J\eta}|}$ and $r_{2}=\sqrt{|\frac{J+t_{0}-J\eta}{J+t_{0}+J\eta}|}$. This behavior differs markedly from the rung-coupling case \cite{li2020critical,yokomizo2021scaling}, where $E^{o}$ must remain on the real axis (App.‌~\ref{c}) and its GBZ forms a single circle with radius $\sqrt{|\frac{1-\eta}{1+\eta}|}$.
	Evidently, the skin decay length is not unique, even in the
	$N\rightarrow\infty$ limit. Adjusting $t_{0}$ can alter the degree of the skin localization, resulting in the appearance of Bloch waves in the strong coupling limit ($r_{1}=r_{2}=1$).
	We will soon demonstrate that our cross-coupling can produce two distinctive types of bulk mode and reveal their novel BBC. Consider the eigenstate ansatz
	
	\begin{equation}\label{44}
	\begin{bmatrix}\mu_{j}\\\nu_{j} \\\end{bmatrix}=\beta_{1}^{j}\begin{bmatrix}\mu^{(1)}\\\nu^{(1)} \\\end{bmatrix}+\beta_{2}^{j}\begin{bmatrix}\mu^{(2)}\\\nu^{(2)} \\\end{bmatrix}+\beta_{3}^{j}\begin{bmatrix}\mu^{(3)}\\\nu^{(3)} \\\end{bmatrix}+\beta_{4}^{j}\begin{bmatrix}\mu^{(4)}\\\nu^{(4)} \\\end{bmatrix}
	\end{equation}
	again. The sublattice symmetry balances the relative weights within the unit cells:
	$\nu^{(2,3)}/\mu^{(2,3)}=-\nu^{(1,4)}/\mu^{(1,4)}=1$, so that  Eq.~(\ref{44}) simplifies to
	
	\begin{equation}\label{45}
	\begin{bmatrix}\mu_{j}\\\nu_{j} \\\end{bmatrix}=\begin{bmatrix}\beta_{1}^{j}\mu^{(1)}+\beta_{2}^{j}\mu^{(2)}+\beta_{3}^{j}\mu^{(3)}+\beta_{4}^{j}\mu^{(4)}\\-\beta_{1}^{j}\mu^{(1)}+\beta_{2}^{j}\mu^{(2)}+\beta_{3}^{j}\mu^{(3)}-\beta_{4}^{j}\mu^{(4)}\\\end{bmatrix}.
	\end{equation}
	Open boundary conditions require:
	\begin{equation}\label{46}
	\mu^{(2)}+\mu^{(3)}=\mu^{(1)}+\mu^{(4)}=0,\,
	\bigg |\frac{\mu^{(4)}}{\mu^{(2)}}\bigg|=\bigg|\frac{\beta_{3}^{N+1}-\beta_{2}^{N+1}}{\beta_{4}^{N+1}-\beta_{1}^{N+1}}\bigg|.
	\end{equation}
	The absolute ratio $|\mu^{(4)}/\mu^{(2)}|$ in Eq.~(\ref{46}), which determines the dominant pair of
	eigenmodes, indicates multimode competition that fragments the GBZ \cite{meng2025generalized}. The
	pronounced discrepancy between the PBC and OBC energy spectra in Fig.~\ref{fig8} clearly signals the re-emergence of the NHSE. Analogous to the well-studied single-band NH chain, a sharp phase transition associated with the NHSE occurs at $\eta=0$ \cite{gong2018topological}.
	Without loss of generality, we assume  $\eta>0$ in the following discussion.
	
	When $J<t_{0}$ making $r_{1}>1>r_{2}$, the entire OBC energy band
	$E^{o}_{+}$  is encircled counterclockwise by a closed PBC loop  evolving as  $E^{p}_{+}(k=-\frac{\pi}{2})\rightarrow E^{p}_{+}(k=\frac{\pi}{2})= E^{p}_{-}(k=\frac{\pi}{2})\rightarrow E^{p}_{-}(k=\frac{3\pi}{2})$; while  $E^{o}_{-}$ is encircled clockwise by the rest of  $E^{p}_{\pm}(k)$ [Fig.~\ref{fig8} (a)]. Invoking the hybrid spectral winding, we obtain
	\begin{equation}\label{47}
	\mathcal{W}^{\pm}_{hs}=
	\int^{\frac{\pi}{2}}_{-\frac{\pi}{2}} \partial_{k}Arg[E^{p}_{\pm}(k)]\frac{dk}{2\pi}+\int^{\frac{3\pi}{2}}_{\frac{\pi}{2}} \partial_{k}Arg[E^{p}_{\mp}(k)]\frac{dk}{2\pi}=\pm1,
	\end{equation}
	where ``$+1$''(``$-1$'') predicts left (right)-localized skin modes. Such non-Hermitian BBC is not strictly restricted to the special case of $\delta=0$, yet this parameter point lends itself far more naturally to a straightforward analytical treatment. For any $E^{o}_{+j}$, the relationship $|\beta_{4}|>1>|\beta_{2}|=|\beta_{3}|>|\beta_{1}|$ ($\beta_{1,4}$  is actually real) will result in $|\mu^{(4 )}/\mu^{(2)}|\approx 0$ according to Eq.~(\ref{46}). This skin mode dominated by $\beta_{2,3}$ is exponentially localized at the left end [inset in the lower right of Fig.~\ref{fig8} (a)], consistent with the conventional GBZ prediction.  For any $E^{o}_{-j}$, Eq.~(\ref{41}) implies that $\beta_{1,4}$ and $\beta_{2,3}$ occupy different GBZ circles with radii $r_{1}$ and $r_{2}$, respectively, definitely violating the standard continuous band condition. However, $\beta_{1,4}=\frac{\sqrt{d_{-}}}{J-t_{0}+J\eta}e^{\pm\frac{ij\pi}{N+1}}$
	leads to $\mu^{(2)}=0$ in Eq.~(\ref{46}), thereby retaining only the 
	modes squeezed to the right [ inset in the lower left of Fig.~\ref{fig8} (a)]. To note,
	this unconventional  bipolar NHSE does not originate from self-intersections of a single PBC band, and the OBC spectra of skin modes squeezed in opposite directions  overlap, thus no so-called NHSE edges as mentioned in Ref.\cite{zeng2022non}.
	Another anomalous feature appears at $J=t_{0}$, where pure imaginary $E^{o}_{-j}=2iJ\eta \cos (\frac{\pi j}{N+1})$ overlape with PBC spectra ($\mathcal{W}_{hs}^{-}=0$),
	forming a Bloch line in the complex plane \cite{song2019non}. From Eq.~(\ref{46}), only the Bloch-wave component with $\beta_{1,4}=ie^{\pm\frac{ ij\pi}{N+1}}$ remains, as shown in the lower-left  inset of Fig.~\ref{fig8} (b). When $J>t_{0}$,  neither a clockwise PBC energy loop nor any anomalous modes exist.

	\subsection{Topological boundary states beyond the hidden chiral symmetry}
	
	Under the pseudo-inversion symmetry with $\eta\neq0$, $\mathcal{\delta}\neq0$ and $\gamma=0$, the recursion equation $\bm{P}^{\prime}|\psi_{j+1}\rangle+\bm{Q}^{\prime}|\psi_{j-1}\rangle=E^{o}|\psi_{j}\rangle$, where
	\begin{equation}\label{48}
	\begin{aligned}
	&\bm{P}^{\prime}=\begin{bmatrix}J(1+\eta)&t_{2} \\t_{1} &J(1+\eta)\\\end{bmatrix},\,\,\bm{Q}^{\prime}=\begin{bmatrix}J(1-\eta)&t_{1} \\t_{2} &J(1-\eta)\\\end{bmatrix},\\
	\end{aligned}
	\end{equation}
	yields two eigenvalues of the transfer matrix $\bm{T}^{\prime}=-\bm{P}^{\prime-1}\bm{Q}^{\prime}$ satisfying the product relation $\mathcal{Z}^{\prime}_{+}\mathcal{Z}^{\prime}_{-}=1-\frac{4\eta J^{2}}{t_{0}^{2}(\delta^{2}-1)+J^{2}(1+\eta)^{2}}$. At the critical point $J=t_{0}$, only one of the two relations
 $\mathcal{Z}^{\prime}_{+}=1$ and $\mathcal{Z}^{\prime}_{-}=1$ holds.
	 In the regime where $J>t_{0}$, $|\mathcal{Z}^{\prime}_{+}|$ and $|\mathcal{Z}^{\prime}_{-}|$ lie on the same side of unity, forcing the two zero-energy modes to localize at the same boundary. Since one of the eigenvalues $\mathcal{Z}^{\prime}_{\pm}$ crosses the unit circle   (i.e., $|\mathcal{Z}^{\prime}_{\pm}|=1$) at $J=t_{0}$ as $J$ decreases, the localization behaviors in the $J<t_{0}$ regime  exhibit distinct characteristics: for odd $N$,  two zero modes are exponentially localized at opposite ends, whereas for even $N$, the weight ratio between the two modes, dictated by the boundary conditions, determines which end the two zero modes localize at. In particular, when  $|\delta|=|\eta|=1$, $E^{o}$ merge into three eigenvalues: $0$ and $\pm2 t_{0}$, thereby giving rise to high-order OBC EPs.  Due to the spontaneous breaking of the hidden chiral symmetry, topological edge states can no longer be characterized  by the $\mathbb{Z}_{2}$ invariant $\mathcal{D}$ or its extended variant.

\section{conclusion}\label{sec.6}
	In this work, we have investigated the impact of lattice symmetries on the OBC bulk and zero-energy states of a non-Hermitian ladder from  the perspective of the BBC. 

Regarding the bulk states, when both the $\mathcal{P}$ and $\mathcal{T}$ symmetries are preserved (Sec.~\ref{sec.3}), the $\mathcal{P}$-symmetric pairs of non-Bloch factors $\beta_{i}^{\pm 1}$ eradicate exponential skin modes but allow for SF skin modes. The presence of the latter has been  independently
	confirmed by the TI. Under the reverse nonreciprocity $\eta_{a}=-\eta_{b}$, the gain-loss breaks the $\mathcal{P}$ symmetry yet preserves the composite $\mathcal{P}\mathcal{T}$ symmetry (Sec.~\ref{sec.4}). The subsequent 
	$\mathcal{P}\mathcal{T}$ symmetry breaking triggers exponential skin modes due to the partial deformation of the BZ. More precisely, only the spectral portion where the $\mathcal{P}\mathcal{T}$ symmetry is broken induce the symmetric bipolar NHSE. This delocalization-localization transition invariably coincides with the breaking of $\mathcal{P}\mathcal{T}$ symmetry identified by a trivial AWN. Upon restoring the sublattice symmetry (Sec.~\ref{sec.5}), the continuous band condition of the GBZ is no longer valid. Through an OBC spectrum analysis that determines two distinct sub-GBZ loops, we have explained the anomalous bipolar NHSE and the unconventional Bloch line, both of which are not caused by the intersection of GBZ and BZ as in a deformed non-Hermitian SSH model \cite{song2019non}, but rather by GBZ fragmentation. The hybrid complex energy index $\mathcal{W}^{\pm}_{hs}$ has illuminated  their topological origin. Notably, the extended mode with maximum imaginary energy in the Bloch line may facilitate broad-area laser emission without requiring precise modulation of nonreciprocal intensity \cite{longhi2018non}.
	
Turn to the zero-energy modes. In the case of separate $\mathcal{P}$ symmetry  (Sec.~\ref{sec.3}), for any odd ladder length $N$, degenerate zero-energy modes persist and remain localized at the boundary exclusively in regimes where $\mathcal{D}=\pm1$. In contrast,  for even $N$, zero-energy modes are absent when $\mathcal{D}\neq\pm 1$; however, a pair of zero-energy edge modes emerges in $\mathcal{D}=\pm 1$  regimes as the system size increases. This even-odd effect distinguishes our model from the CNHSE-SSH model with non-reciprocal cross-coupling \cite{li2020critical}, in which the zero-energy modes are also edge modes characterized by the Berry phase of a single non-Bloch band, a feature originating from the absence of $\sigma_{x}$ in the Hamiltonian. In our case, topological localization  manifests solely when $\mathcal{D}=\pm1$, regardless of whether the ladder length is odd or even. The capability of the Bloch bulk $\mathbb{Z}_{2}$ invariant $\mathcal{D}$ to simultaneously characterize zero-energy edge modes in both $\mathcal{P}\mathcal{T}$ BP and UBP stems from the fact that the splitting (merging) of OBC spectra, corresponding to the isolation (collision) of PBC spectra, occurs strictly within the real  spectral segment. The proliferation of localized zero modes at the EP is anticipated to offer unique applications in flatband engineering \cite{esparza2025exceptional}. Furthermore,  $\mathcal{P}\mathcal{T}$-protected gain-loss does not violate the hidden chiral symmetry, hence the redefined Bloch $\mathbb{Z}_{2}$  invariant remains effective even in the presence of NHSE, as discussed in Sec.~\ref{sec.4}.

The tunability of skin and topological edge modes via symmetry switching in a single system is highly desirable for the thorough understanding  of non-Hermitian phase transitions, and offers great flexibility in the application and design of non-Hermitian optoelectronic devices \cite{jiang2023reciprocating}, directional amplifiers \cite{PhysRevB.103.L241408} and symmetry-protected sensors \cite{parto2025enhanced}. Our model provides a starting point to explore two-dimensional non-Hermitian topological systems with analogous  higher-order properties. Building upon the comprehensive physical insights provided by this work, we anticipate the experimental verification of these exotic phenomena across diverse classical and quantum platforms.
	
	\section*{Acknowledgments} 
	This work was supported by Yunnan Ten Thousand Talents Plan Young and Elite Talents Project (Grant Number:YNWR-QNBJ-2018-121).

	\appendix
	
	\section{}\label{a}
	Consider a Hamiltonian $H$ represented as an $N\times N$ matrix. The IPR in a preferential basis  $\{|j\rangle\}^{N}_{j=1}$ defined as \cite{misguich2016inverse} 
	\begin{equation}\label{48}
	IPR[|\psi\rangle]=\frac{\sum_{j=1}^{N}|\langle j|\psi\rangle|^{4}}{(\sum_{j=1}^{N}|\langle j|\psi\rangle|^{2})^{2}}
	\end{equation}
	is widely  employed to measure the spread of an eigenstate $|\psi\rangle$ of $H$. The maximum value $IPR[|\psi\rangle]=1$ signifies that $|\psi\rangle$ is strictly localized on a single site $|j\rangle$. Conversely, the minimum value $IPR[|\psi\rangle]=1/N$ corresponds to a delocalized limit whereby $|\psi\rangle$ uniformly populates all $|j\rangle$. However, the
	IPR fails to accurately capture the nature of SF localization. For instance, in a unidirectional chain with generalized boundary conditions described by $H=\sum_{J=1}^{N-1}|j\rangle\langle j+1|+\zeta|N\rangle\langle 1|(1\geq\zeta\geq0)$ \cite{guo2021exact, zhang2025scale}, the eigenstates
	\begin{equation}\label{49}
	|\psi_{m}\rangle=[1,\zeta^{\frac{1}{N}}e^{i\vartheta_{m}},\ldots,\zeta^{\frac{N-1}{N}}e^{i(N-1)\vartheta_{m}}]^{T}
	\end{equation}
	($\vartheta_{m}=\frac{2m\pi}{N},m=1,2,\ldots N$) represent typical SF modes. Nevertheless, their
	asymptotic IPR vanishes in the thermodynamic limit $\lim _{N\rightarrow\infty}IPR[|\psi_{m}\rangle]=0$, as
	shown by the exact expression
	\begin{equation}\label{50}
	IPR[|\psi_{m}\rangle]=\frac{\sum_{j=1}^{N}\zeta^{\frac{4(j-1)}{N}}}{(\sum_{j=1}^{N}\zeta^{\frac{2(j-1)}{N}})^{2}}=\dfrac{(\zeta^{2}+1)(\zeta^{\frac{2}{N}}-1)}{(\zeta^{2}-1)(\zeta^{\frac{2}{N}}+1)}.
	\end{equation}
	
	\section{}\label{b}
	Although  the Bloch Hamiltonian $\mathcal{H}(k)$ in Eq.~(\ref{2}) lacks conventional chiral symmetry, it exhibits a hidden chiral symmetry given by \cite{Li_2015}
	\begin{equation}\label{51}
	\sigma_{x}\mathcal{H}(k)\sigma_{x}=-\mathcal{H}(\pi-k).
	\end{equation}
	In the OBC case, the real-space Hamiltonian can be recast into the block matrix form 
	
	\begin{equation}\label{52}
	H=\begin{bmatrix}H_{HN}&H_{I}\\H_{I}^{T}&H_{HN}^{T}\\\end{bmatrix},
	\end{equation}
	where the basis is ordered as $[1a,2a,\cdots,Na,1b,2b,\cdots,Nb]$. Here,
	\begin{equation}\label{53}
	H_{HN}=\begin{bmatrix}0&J(1+\eta)&0&\cdots\\J(1-\eta)&0&J(1+\eta)&\cdots\\0&J(1-\eta)&0&\cdots\\\cdots&\cdots&\cdots&\ddots\\\end{bmatrix}
	\end{equation}
	represents the Hamiltonian of a single NH chain, and
	
	\begin{equation}\label{54}
	H_{I}=\begin{bmatrix}0&t_{0}(1+\delta)&0&\cdots\\t_{0}(1-\delta)&0&t_{0}(1+\delta)&\cdots\\0&t_{0}(1-\delta)&0&\cdots\\\cdots&\cdots&\cdots&\ddots\\\end{bmatrix}
	\end{equation}
	describes the inter-chain coupling. The chiral
	symmetry of the total Hamiltonian
	\begin{equation}\label{55}
	\bm{C}H\bm{C}^{-1}=-H
	\end{equation}
can be explicitly verified by constructing the hidden chiral operator
	\begin{equation}\label{56}
	\bm{C}=\begin{bmatrix}0&\Upsilon\\\Upsilon&0\\\end{bmatrix},
	\end{equation}
	where the unitary matrix $\bm{C}$ satisfies $\bm{C}^{2}=-1$. The $\Upsilon$ is a $N\times N$ matrix with non-zero elements  $\Upsilon_{n,N-n+1}=i^{N}(-1)^{n+1}$. Consequently, for every eigenstate $|\mathcal{R}_{j}\rangle$ satisfying $H|\mathcal{R}_{j}\rangle=E^{o}_{j}|\mathcal{R}_{j}\rangle$,  there exists a partner state    $\bm{C}^{-1}|\mathcal{R}_{j}\rangle$ such that $H\bm{C}^{-1}|\mathcal{R}_{j}\rangle=-E^{o}_{j}\bm{C}^{-1}|\mathcal{R}_{j}\rangle$. At zero energy $E^{o}=0$, $|\mathcal{R}^{(\prime)}_{\pm}\rangle=\bm{C}|\mathcal{R}^{(\prime)}_{\pm}\rangle$ implies that the zero modes are topologically protected by this hidden chiral symmetry. In realistic systems, imperfections and constraints may prevent the strict preservation of this symmetry, rendering $|\mathcal{R}^{(\prime)}_{\pm}\rangle$ no longer an exact zero mode. However, as long as the mode energy remains well-separated from the bulk bands, the topological features are robustly inherited, as corroborated by numerous experiments.

	\section{}\label{c}

	In this section, we address the eigenvalue problem of  $H(\delta=\gamma=0,\eta_{a}=\eta_{b}=\eta)$ under OBC. Evidently, $E^{o}_{\pm j}$ in Eq.~(\ref{43}) should be the roots 
	of the characteristic dispersion polynomial
	\begin{equation}\label{57}
	\begin{aligned}
	&&\det [H-E^{o}]=\det\begin{bmatrix}H_{HN}-E^{o}&H_{I}^{0}\\H_{I}^{0}&H_{HN}-E^{o}\\\end{bmatrix}=\,\,\,\,\,\\\
	&&\det [H_{HN}-E^{o}+H_{I}^{0}]\det [H_{HN}-E^{o}-H_{I}^{0}]=0,\\
	\end{aligned}
	\end{equation}
	with $H_{I}^{0}\equiv H_{I}(\delta=0)$. Eq.~(\ref{57}) decouples the spectrum $E^{o}$ into two branches , which can be  solved exactly via similarity transformations
	\begin{equation}\label{58}
	\bm{S}_{\pm}(H_{HN}-E^{o}\pm H_{I}^{0})\bm{S}_{\pm}^{-1}=\begin{bmatrix}-E^{o}&\sqrt{d_{\pm}}&0&\cdots\\\sqrt{d_{\pm}}&-E^{o}&\sqrt{d_{\pm}}&\cdots\\0&\sqrt{d_{\pm}}&-E^{o}&\cdots\\\cdots&\cdots&\cdots&\ddots\\\end{bmatrix},
	\end{equation}
	where $\bm{S}_{\pm}$ are the diagonal matrices with  entries $\{(\frac{J\pm t_{0}+J\eta}{J\pm t_{0}-J\eta})^{\frac{1-N}{4}},(\frac{J\pm t_{0}+J\eta}{J\pm t_{0}-J\eta})^{\frac{3-N}{4}},\cdots,(\frac{J\pm t_{0}+J\eta}{J\pm t_{0}-J\eta})^{\frac{N-1}{4}}\}$, because the matrix in Eq.~(\ref{58}) can be diagonalized following the same mathematical procedure established for Hermitian single-mode waveguides \cite{rai2008transport}. In the presence of rung coupling, $t_{0}$ appears solely in the diagonal
	terms of the interaction matrix. Thus,  the eigenvalues are readily obtained as $E^{o}_{\pm j}=2J\sqrt{1-\eta^{2}}\cos (\frac{j\pi}{N+1})\pm t_{0}$.

	\bibliographystyle{apsrev4-2}
	\bibliography{Maintextreference}

\end{document}